\documentclass[longauth]{aa}

\makeatletter
\renewcommand*\maketitle{%
  \thispagestyle{firstpage}
\begingroup
    \if@wideboxfn
    \setlength\bibindent{1.4\parindent}
    \else
    \setlength\bibindent{\parindent}
    \fi
    \renewcommand*\thefootnote{\@fnsymbol\c@footnote}%
    \renewcommand\@makefntext[1]{%
    \ifaa@longfn\hsize\textwidth\fi
    \noindent
    \hb@xt@\bibindent{\hss\@makefnmark\enspace}##1}
  \ifaa@twocolumn
  \begingroup
    \begin{aa@strip}
          \aa@maketitle
    \end{aa@strip}
    \@thanks
  \endgroup
  \else
    \begingroup
      \let\thanks\footnote
      \aa@maketitle
    \endgroup
  \fi
\endgroup
  \setcounter{footnote}{0}%
}
\makeatother

\usepackage{natbib,twoopt}
\usepackage{xcolor}
\usepackage{wasysym}
\usepackage[breaklinks=true]{hyperref} 
\usepackage{graphicx}
\usepackage{txfonts}
\usepackage{textcomp}
\usepackage{array}
\usepackage{gensymb}
\usepackage{listings}
\usepackage{ulem}
\usepackage{hyperref}

\newcommand{\orcit}[1]{\protect\href{https://orcid.org/#1}{\protect\includegraphics[width=8pt]{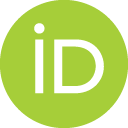}}}

\bibpunct{(}{)}{;}{a}{}{,}             
\makeatletter
\newcommandtwoopt{\citeads}[3][][]{\href{http://adsabs.harvard.edu/abs/#3}%
  {\def\hyper@linkstart##1##2{}%
    \let\hyper@linkend\@empty\citealp[#1][#2]{#3}}}
\newcommandtwoopt{\citepads}[3][][]{\href{http://adsabs.harvard.edu/abs/#3}%
  {\def\hyper@linkstart##1##2{}%
    \let\hyper@linkend\@empty\citep[#1][#2]{#3}}}
\newcommandtwoopt{\citetads}[3][][]{\href{http://adsabs.harvard.edu/abs/#3}%
  {\def\hyper@linkstart##1##2{}%
    \let\hyper@linkend\@empty\citet[#1][#2]{#3}}}
\newcommandtwoopt{\citeyearads}[3][][]%
{\href{http://adsabs.harvard.edu/abs/#3}
  {\def\hyper@linkstart##1##2{}%
    \let\hyper@linkend\@empty\citeyear[#1][#2]{#3}}}
\renewcommand*\aa@pageof{, page \thepage{} of \pageref*{LastPage}}
\makeatother
\hypersetup{colorlinks=true,linkcolor=blue,citecolor=blue,urlcolor=blue}

\providecommand{\orcit}[1]{\protect\href{https://orcid.org/#1}{\protect\includegraphics[width=8pt]{orcid.png}}}

\def\deg{\ensuremath{^\circ}}

\def\arcsec{\ensuremath{''}}

\providecommand{\teff}{\ensuremath{{\mathrm{T_{eff}}}}\xspace}

\providecommand{\logg}{\ensuremath{\log\,g}\xspace}

\providecommand{\loggrav}{\ensuremath{\log\,g}\xspace}

\providecommand{\vbroad}{\ensuremath{v_\mathrm{broad}}\xspace}

\providecommand{\bprp}{BP/RP\xspace}

\providecommand{\ag}{\ensuremath{A_G}\xspace}
\providecommand{\bpminrp}{\ensuremath{(G_\mathrm{BP}-G_\mathrm{RP})}\xspace}

\providecommand{\bpmag}{\ensuremath{G_\mathrm{BP}}\xspace}
\providecommand{\rpmag}{\ensuremath{G_\mathrm{RP}}\xspace}

\providecommand\gmag{\ensuremath{G}\xspace}
\providecommand\gbp{\ensuremath{G_\mathrm{BP}}\xspace}
\providecommand\grp{\ensuremath{G_\mathrm{RP}}\xspace}
\providecommand\grvs{\ensuremath{G_\mathrm{RVS}}\xspace}

\providecommand{\kpc}{\ensuremath{\,\rm kpc}\xspace}

\providecommand{\ang}{\ensuremath{\,\text{\AA}}\xspace}
\providecommand{\mas}{\ensuremath{\,\mathrm{mas}}\xspace}
\providecommand{\kms}{\ensuremath{\textrm{km\,s}^{-1}}}
\providecommand{\maspyr}{\ensuremath{\textrm{mas\,yr}^{-1}}}

\providecommand\muasyr{\ensuremath{\mu\text{as\,yr}^{-1}}}
\providecommand\kms{\ensuremath{\text{\,km\,s}^{-1}}}

\providecommand\muas{\ensuremath{\mu\text{as}}\xspace}

\providecommand{\modulename}[1]{#1\xspace}

\providecommand{\gspphot}{\modulename{GSP-Phot}}

\providecommand{\gspspec}{\modulename{GSP-Spec}}

\newcommand\gaia{\textit{Gaia}\xspace}
\providecommand\hip{\textsc{Hipparcos\xspace}}
\providecommand\tyc{\textit{Tycho\xspace}}
\providecommand\tyctwo{\textit{Tycho}-2\xspace}

\providecommand\gdrtwo{\textit{Gaia}~DR2\xspace}
\providecommand\gedrthree{\textit{Gaia}~EDR3\xspace}
\providecommand\gdrthree{\textit{Gaia}~DR3\xspace}
\providecommand\gdr[1]{\textit{Gaia}~DR#1\xspace}
\providecommand\gedr[1]{\textit{Gaia}~EDR#1\xspace}

\definecolor{dkgreen}{rgb}{0,0.6,0}
\definecolor{gray}{rgb}{0.5,0.5,0.5}
\definecolor{mauve}{rgb}{0.58,0,0.82}
\definecolor{golden}{rgb}{0.86,0.65,0.01}

\providecommand\figref[1]{Fig.~\ref{#1}}

\providecommand\figrefalt[1]{Figure~\ref{#1}}

\providecommand\tabref[1]{Table~\ref{#1}}

\providecommand{\linktodoc}{https://gea.esac.esa.int/archive/documentation/GDR3}

\providecommand{\linktotableap}[1]{\url{\linktodoc/Gaia_archive/chap_datamodel/sec_dm_astrophysical_parameter_tables/ssec_dm_#1.html}} 

\providecommand{\gstable}[1]{\url{\linktodoc/Gaia_archive/chap_datamodel/sec_dm_main_source_catalogue/ssec_dm_#1.html}}
\providecommand{\varitable}[1]{\url{\linktodoc/Gaia_archive/chap_datamodel/sec_dm_variability_tables/ssec_dm_#1.html}}
\providecommand{\phottable}[1]{\url{\linktodoc/Gaia_archive/chap_datamodel/sec_dm_spectroscopic_tables/ssec_dm_#1.html}}
\providecommand{\nsstable}[1]{\url{\linktodoc/Gaia_archive/chap_datamodel/sec_dm_non--single_stars_tables/ssec_dm_#1.html}} 
\providecommand{\ssotable}[1]{\url{\linktodoc/Gaia_archive/chap_datamodel/sec_dm_solar_system_object_tables/ssec_dm_#1.html}} 
\providecommand{\pvptable}[1]{\url{\linktodoc/Gaia_archive/chap_datamodel/sec_dm_performance_verification/ssec_dm_#1.html}}
\providecommand{\extragal}[1]{\url{\linktodoc/Gaia_archive/chap_datamodel/sec_dm_extra--galactic_tables/ssec_dm_#1.html}}

\providecommand{\linktotable}[1]{\url{\linktodoc/Gaia\_archive/chap\_datamodel/sec\_dm\_main\_tables/ssec\_dm\_#1.html}} 

\providecommand{\linkfig}[1]{\href{\linktodoc/Data_analysis/chap_cu8par/#1}{see table\xspace}}

\makeatletter
\DeclareRobustCommand*{\fieldName}[1]{%
  \begingroup\@fieldName\scantokens{\texttt{\small {#1}}\noexpand}\endgroup}
\begingroup\lccode`\~=`\_\relax
   \lowercase{\endgroup\def\@fieldName{\catcode`\_=\active \let~\_}}
\makeatother

\newcommand\gdrthreetotal{\ensuremath{1\,811\,709\,771}\xspace}
\newcommand\gdrthreefiveptot{\ensuremath{585\,416\,709}\xspace}
\newcommand\gdrthreesixptot{\ensuremath{882\,328\,109}\xspace}
\newcommand\gdrthreetwoptot{\ensuremath{343\,964\,953}\xspace}
\newcommand\gdrthreegcrftot{\ensuremath{1\,614\,173}\xspace}
\newcommand\gdrthreeicrfused{\ensuremath{2007}\xspace}
\newcommand\gdrthreespinused{\ensuremath{428\,034}\xspace}

\newcommand\gdrthreewithgtot{\ensuremath{1\,806\,254\,432}\xspace}
\newcommand\gdrthreewithbptot{\ensuremath{1\,542\,033\,472}\xspace}
\newcommand\gdrthreewithrptot{\ensuremath{1\,554\,997\,939}\xspace}

\newcommand\gdrthreevradtot{\ensuremath{33\,812\,183}\xspace}
\newcommand\gdrthreegrvstot{\ensuremath{32\,232\,187}\xspace}
\newcommand\gdrthreevbroadtot{\ensuremath{3\,524\,677}\xspace}
\newcommand\gdrthreeepochrvtot{\ensuremath{1898}\xspace}

\newcommand{\gdrthreexpspectratot}{\ensuremath{219\,197\,643}\xspace}
\newcommand{\gdrthreervsspectratot}{\ensuremath{999\,645}\xspace}

\newcommand{\gdrthreevaritot}{\ensuremath{10\,509\,536}\xspace}
\newcommand{\gdrthreevarimlclassified}{\ensuremath{9\,976\,881}\xspace}

\newcommand{\gdrthreephottimeseries}{\ensuremath{11\,754\,237}\xspace}
\newcommand{\gdrthreegaps}{\ensuremath{1\,257\,319}\xspace}
\newcommand{\gdrthreevaricepheids}{\ensuremath{15\,021}\xspace}
\newcommand{\gdrthreevaricompact}{\ensuremath{6\,306}\xspace}
\newcommand{\gdrthreevarieclipsing}{\ensuremath{2\,184\,477}\xspace}
\newcommand{\gdrthreevarilpv}{\ensuremath{1\,720\,588}\xspace}
\newcommand{\gdrthreevarimulens}{\ensuremath{363}\xspace}
\newcommand{\gdrthreevariplanets}{\ensuremath{214}\xspace}
\newcommand{\gdrthreevarirrl}{\ensuremath{271\,779}\xspace}
\newcommand{\gdrthreevarishortts}{\ensuremath{471\,679}\xspace}
\newcommand{\gdrthreevarisolarlike}{\ensuremath{474\,026}\xspace}
\newcommand{\gdrthreevariums}{\ensuremath{54\,476}\xspace}
\newcommand{\gdrthreevariagn}{\ensuremath{872\,228}\xspace}

\newcommand{\gdrthreedsc}{\ensuremath{1\,590\,760\,469}\xspace}
\newcommand{\gdrthreeapsfromxp}{\ensuremath{470\,759\,263}\xspace}
\newcommand{\gdrthreeapsbinary}{\ensuremath{348\,711\,151}\xspace}

\newcommand{\gdrthreemcmcmsc}{\ensuremath{348\,711\,151}\xspace}
\newcommand{\gdrthreemcmcgspphot}{\ensuremath{449\,297\,716}\xspace}
\newcommand{\gdrthreeevolpars}{\ensuremath{128\,611\,111}\xspace}
\newcommand{\gdrthreexpels}{\ensuremath{57\,511}\xspace}
\newcommand{\gdrthreexpspt}{\ensuremath{217\,982\,837}\xspace}
\newcommand{\gdrthreexphotstars}{\ensuremath{2\,382\,015}\xspace}
\newcommand{\gdrthreexpucd}{\ensuremath{94\,158}\xspace}
\newcommand{\gdrthreexpcoolactive}{\ensuremath{1\,349\,499}\xspace}
\newcommand{\gdrthreexphalpha}{\ensuremath{235\,384\,119}\xspace}

\newcommand{\gdrthreeapsfromrvs}{\ensuremath{5\,591\,594}\xspace}
\newcommand{\gdrthreeabunfromrvs}{\ensuremath{2\,513\,593}\xspace}
\newcommand{\gdrthreedibs}{\ensuremath{472\,584}\xspace}

\newcommand{\gdrthreenss}{\ensuremath{813\,687}\xspace}
\newcommand{\gdrthreenssacceleration}{\ensuremath{338\,215}\xspace}

\newcommand{\gdrthreenssorbitalastromcomb}{\ensuremath{169\,227}\xspace}

\newcommand{\gdrthreenssorbitalspectrocomb}{\ensuremath{220\,372}\xspace}
\newcommand{\gdrthreensstrendspectro}{\ensuremath{56\,808}\xspace}
\newcommand{\gdrthreensseclipsing}{\ensuremath{87\,073}\xspace}

\newcommand{\gdrthreeqsocandidates}{\ensuremath{6\,649\,162}\xspace}
\newcommand{\gdrthreeqsoredshifts}{\ensuremath{6\,375\,063}\xspace}
\newcommand{\gdrthreeqsohosts}{\ensuremath{64\,498}\xspace}
\newcommand{\gdrthreeqsohostprofile}{\ensuremath{15\,867}\xspace}

\newcommand{\gdrthreegalaxycandidates}{\ensuremath{4\,842\,342}\xspace}
\newcommand{\gdrthreegalaxyredshifts}{\ensuremath{1\,367\,153}\xspace}
\newcommand{\gdrthreegalaxyprofile}{\ensuremath{914\,837}\xspace}

\newcommand{\gdrthreesso}{\ensuremath{158\,152}\xspace}
\newcommand{\gdrthreessoreflectance}{\ensuremath{60\,518}\xspace}

\newcommand{\gdrthreealerts}{\ensuremath{2612}\xspace}

\begin{document}

\title{{\gaia} Data Release 3:}
\subtitle{Summary of the content and survey properties}
\titlerunning{{\gaia} Data Release 3: Summary}
\authorrunning{Gaia Collaboration}

\author{
{\it Gaia} Collaboration
\and         A.~                     Vallenari\orcit{0000-0003-0014-519X}\inst{\ref{inst:0001}}
\and     A.G.A.~                         Brown\orcit{0000-0002-7419-9679}\inst{\ref{inst:0002}}
\and         T.~                        Prusti\orcit{0000-0003-3120-7867}\inst{\ref{inst:0003}}
\and     J.H.J.~                    de Bruijne\orcit{0000-0001-6459-8599}\inst{\ref{inst:0003}}
\and         F.~                        Arenou\orcit{0000-0003-2837-3899}\inst{\ref{inst:0005}}
\and         C.~                     Babusiaux\orcit{0000-0002-7631-348X}\inst{\ref{inst:0006},\ref{inst:0005}}
\and         M.~                      Biermann\inst{\ref{inst:0008}}
\and       O.L.~                       Creevey\orcit{0000-0003-1853-6631}\inst{\ref{inst:0009}}
\and         C.~                     Ducourant\orcit{0000-0003-4843-8979}\inst{\ref{inst:0010}}
\and       D.W.~                         Evans\orcit{0000-0002-6685-5998}\inst{\ref{inst:0011}}
\and         L.~                          Eyer\orcit{0000-0002-0182-8040}\inst{\ref{inst:0012}}
\and         R.~                        Guerra\orcit{0000-0002-9850-8982}\inst{\ref{inst:0013}}
\and         A.~                        Hutton\inst{\ref{inst:0014}}
\and         C.~                         Jordi\orcit{0000-0001-5495-9602}\inst{\ref{inst:0015}}
\and       S.A.~                       Klioner\orcit{0000-0003-4682-7831}\inst{\ref{inst:0016}}
\and       U.L.~                       Lammers\orcit{0000-0001-8309-3801}\inst{\ref{inst:0013}}
\and         L.~                     Lindegren\orcit{0000-0002-5443-3026}\inst{\ref{inst:0018}}
\and         X.~                          Luri\orcit{0000-0001-5428-9397}\inst{\ref{inst:0015}}
\and         F.~                       Mignard\inst{\ref{inst:0009}}
\and         C.~                         Panem\inst{\ref{inst:0021}}
\and         D.~            Pourbaix$^\dagger$\orcit{0000-0002-3020-1837}\inst{\ref{inst:0022},\ref{inst:0023}}
\and         S.~                       Randich\orcit{0000-0003-2438-0899}\inst{\ref{inst:0024}}
\and         P.~                    Sartoretti\inst{\ref{inst:0005}}
\and         C.~                      Soubiran\orcit{0000-0003-3304-8134}\inst{\ref{inst:0010}}
\and         P.~                         Tanga\orcit{0000-0002-2718-997X}\inst{\ref{inst:0009}}
\and       N.A.~                        Walton\orcit{0000-0003-3983-8778}\inst{\ref{inst:0011}}
\and     C.A.L.~                  Bailer-Jones\inst{\ref{inst:0029}}
\and         U.~                       Bastian\orcit{0000-0002-8667-1715}\inst{\ref{inst:0008}}
\and         R.~                       Drimmel\orcit{0000-0002-1777-5502}\inst{\ref{inst:0031}}
\and         F.~                        Jansen\inst{\ref{inst:0032}}
\and         D.~                          Katz\orcit{0000-0001-7986-3164}\inst{\ref{inst:0005}}
\and       M.G.~                      Lattanzi\orcit{0000-0003-0429-7748}\inst{\ref{inst:0031},\ref{inst:0035}}
\and         F.~                   van Leeuwen\inst{\ref{inst:0011}}
\and         J.~                        Bakker\inst{\ref{inst:0013}}
\and         C.~                      Cacciari\orcit{0000-0001-5174-3179}\inst{\ref{inst:0038}}
\and         J.~                 Casta\~{n}eda\orcit{0000-0001-7820-946X}\inst{\ref{inst:0039}}
\and         F.~                     De Angeli\orcit{0000-0003-1879-0488}\inst{\ref{inst:0011}}
\and         C.~                     Fabricius\orcit{0000-0003-2639-1372}\inst{\ref{inst:0015}}
\and         M.~                     Fouesneau\orcit{0000-0001-9256-5516}\inst{\ref{inst:0029}}
\and         Y.~                    Fr\'{e}mat\orcit{0000-0002-4645-6017}\inst{\ref{inst:0043}}
\and         L.~                     Galluccio\orcit{0000-0002-8541-0476}\inst{\ref{inst:0009}}
\and         A.~                      Guerrier\inst{\ref{inst:0021}}
\and         U.~                        Heiter\orcit{0000-0001-6825-1066}\inst{\ref{inst:0046}}
\and         E.~                        Masana\orcit{0000-0002-4819-329X}\inst{\ref{inst:0015}}
\and         R.~                      Messineo\inst{\ref{inst:0048}}
\and         N.~                       Mowlavi\orcit{0000-0003-1578-6993}\inst{\ref{inst:0012}}
\and         C.~                       Nicolas\inst{\ref{inst:0021}}
\and         K.~                  Nienartowicz\orcit{0000-0001-5415-0547}\inst{\ref{inst:0051},\ref{inst:0052}}
\and         F.~                       Pailler\orcit{0000-0002-4834-481X}\inst{\ref{inst:0021}}
\and         P.~                       Panuzzo\orcit{0000-0002-0016-8271}\inst{\ref{inst:0005}}
\and         F.~                        Riclet\inst{\ref{inst:0021}}
\and         W.~                          Roux\orcit{0000-0002-7816-1950}\inst{\ref{inst:0021}}
\and       G.M.~                      Seabroke\orcit{0000-0003-4072-9536}\inst{\ref{inst:0057}}
\and         R.~                         Sordo\orcit{0000-0003-4979-0659}\inst{\ref{inst:0001}}
\and         F.~                  Th\'{e}venin\inst{\ref{inst:0009}}
\and         G.~                  Gracia-Abril\inst{\ref{inst:0060},\ref{inst:0008}}
\and         J.~                       Portell\orcit{0000-0002-8886-8925}\inst{\ref{inst:0015}}
\and         D.~                      Teyssier\orcit{0000-0002-6261-5292}\inst{\ref{inst:0063}}
\and         M.~                       Altmann\orcit{0000-0002-0530-0913}\inst{\ref{inst:0008},\ref{inst:0065}}
\and         R.~                        Andrae\orcit{0000-0001-8006-6365}\inst{\ref{inst:0029}}
\and         M.~                        Audard\orcit{0000-0003-4721-034X}\inst{\ref{inst:0012},\ref{inst:0052}}
\and         I.~                Bellas-Velidis\inst{\ref{inst:0069}}
\and         K.~                        Benson\inst{\ref{inst:0057}}
\and         J.~                      Berthier\orcit{0000-0003-1846-6485}\inst{\ref{inst:0071}}
\and         R.~                        Blomme\orcit{0000-0002-2526-346X}\inst{\ref{inst:0043}}
\and       P.W.~                       Burgess\inst{\ref{inst:0011}}
\and         D.~                      Busonero\orcit{0000-0002-3903-7076}\inst{\ref{inst:0031}}
\and         G.~                         Busso\orcit{0000-0003-0937-9849}\inst{\ref{inst:0011}}
\and         H.~                   C\'{a}novas\orcit{0000-0001-7668-8022}\inst{\ref{inst:0063}}
\and         B.~                         Carry\orcit{0000-0001-5242-3089}\inst{\ref{inst:0009}}
\and         A.~                       Cellino\orcit{0000-0002-6645-334X}\inst{\ref{inst:0031}}
\and         N.~                         Cheek\inst{\ref{inst:0079}}
\and         G.~                    Clementini\orcit{0000-0001-9206-9723}\inst{\ref{inst:0038}}
\and         Y.~                      Damerdji\orcit{0000-0002-3107-4024}\inst{\ref{inst:0081},\ref{inst:0082}}
\and         M.~                      Davidson\inst{\ref{inst:0083}}
\and         P.~                    de Teodoro\inst{\ref{inst:0013}}
\and         M.~              Nu\~{n}ez Campos\inst{\ref{inst:0014}}
\and         L.~                    Delchambre\orcit{0000-0003-2559-408X}\inst{\ref{inst:0081}}
\and         A.~                      Dell'Oro\orcit{0000-0003-1561-9685}\inst{\ref{inst:0024}}
\and         P.~                        Esquej\orcit{0000-0001-8195-628X}\inst{\ref{inst:0088}}
\and         J.~   Fern\'{a}ndez-Hern\'{a}ndez\inst{\ref{inst:0089}}
\and         E.~                        Fraile\inst{\ref{inst:0088}}
\and         D.~                      Garabato\orcit{0000-0002-7133-6623}\inst{\ref{inst:0091}}
\and         P.~              Garc\'{i}a-Lario\orcit{0000-0003-4039-8212}\inst{\ref{inst:0013}}
\and         E.~                        Gosset\inst{\ref{inst:0081},\ref{inst:0023}}
\and         R.~                       Haigron\inst{\ref{inst:0005}}
\and      J.-L.~                     Halbwachs\orcit{0000-0003-2968-6395}\inst{\ref{inst:0096}}
\and       N.C.~                        Hambly\orcit{0000-0002-9901-9064}\inst{\ref{inst:0083}}
\and       D.L.~                      Harrison\orcit{0000-0001-8687-6588}\inst{\ref{inst:0011},\ref{inst:0099}}
\and         J.~                 Hern\'{a}ndez\orcit{0000-0002-0361-4994}\inst{\ref{inst:0013}}
\and         D.~                    Hestroffer\orcit{0000-0003-0472-9459}\inst{\ref{inst:0071}}
\and       S.T.~                       Hodgkin\orcit{0000-0002-5470-3962}\inst{\ref{inst:0011}}
\and         B.~                          Holl\orcit{0000-0001-6220-3266}\inst{\ref{inst:0012},\ref{inst:0052}}
\and         K.~                    Jan{\ss}en\orcit{0000-0002-8163-2493}\inst{\ref{inst:0105}}
\and         G.~          Jevardat de Fombelle\inst{\ref{inst:0012}}
\and         S.~                        Jordan\orcit{0000-0001-6316-6831}\inst{\ref{inst:0008}}
\and         A.~                 Krone-Martins\orcit{0000-0002-2308-6623}\inst{\ref{inst:0108},\ref{inst:0109}}
\and       A.C.~                     Lanzafame\orcit{0000-0002-2697-3607}\inst{\ref{inst:0110},\ref{inst:0111}}
\and         W.~                  L\"{ o}ffler\inst{\ref{inst:0008}}
\and         O.~                       Marchal\orcit{ 0000-0001-7461-892}\inst{\ref{inst:0096}}
\and       P.M.~                       Marrese\orcit{0000-0002-8162-3810}\inst{\ref{inst:0114},\ref{inst:0115}}
\and         A.~                      Moitinho\orcit{0000-0003-0822-5995}\inst{\ref{inst:0108}}
\and         K.~                      Muinonen\orcit{0000-0001-8058-2642}\inst{\ref{inst:0117},\ref{inst:0118}}
\and         P.~                       Osborne\inst{\ref{inst:0011}}
\and         E.~                       Pancino\orcit{0000-0003-0788-5879}\inst{\ref{inst:0024},\ref{inst:0115}}
\and         T.~                       Pauwels\inst{\ref{inst:0043}}
\and         A.~                  Recio-Blanco\orcit{0000-0002-6550-7377}\inst{\ref{inst:0009}}
\and         C.~                     Reyl\'{e}\orcit{0000-0003-2258-2403}\inst{\ref{inst:0124}}
\and         M.~                        Riello\orcit{0000-0002-3134-0935}\inst{\ref{inst:0011}}
\and         L.~                     Rimoldini\orcit{0000-0002-0306-585X}\inst{\ref{inst:0052}}
\and         T.~                      Roegiers\orcit{0000-0002-1231-4440}\inst{\ref{inst:0127}}
\and         J.~                       Rybizki\orcit{0000-0002-0993-6089}\inst{\ref{inst:0029}}
\and       L.M.~                         Sarro\orcit{0000-0002-5622-5191}\inst{\ref{inst:0129}}
\and         C.~                        Siopis\orcit{0000-0002-6267-2924}\inst{\ref{inst:0022}}
\and         M.~                         Smith\inst{\ref{inst:0057}}
\and         A.~                      Sozzetti\orcit{0000-0002-7504-365X}\inst{\ref{inst:0031}}
\and         E.~                       Utrilla\inst{\ref{inst:0014}}
\and         M.~                   van Leeuwen\orcit{0000-0001-9698-2392}\inst{\ref{inst:0011}}
\and         U.~                         Abbas\orcit{0000-0002-5076-766X}\inst{\ref{inst:0031}}
\and         P.~               \'{A}brah\'{a}m\orcit{0000-0001-6015-646X}\inst{\ref{inst:0136},\ref{inst:0137}}
\and         A.~                Abreu Aramburu\inst{\ref{inst:0089}}
\and         C.~                         Aerts\orcit{0000-0003-1822-7126}\inst{\ref{inst:0139},\ref{inst:0140},\ref{inst:0029}}
\and       J.J.~                        Aguado\inst{\ref{inst:0129}}
\and         M.~                          Ajaj\inst{\ref{inst:0005}}
\and         F.~                 Aldea-Montero\inst{\ref{inst:0013}}
\and         G.~                     Altavilla\orcit{0000-0002-9934-1352}\inst{\ref{inst:0114},\ref{inst:0115}}
\and       M.A.~                   \'{A}lvarez\orcit{0000-0002-6786-2620}\inst{\ref{inst:0091}}
\and         J.~                         Alves\orcit{0000-0002-4355-0921}\inst{\ref{inst:0148}}
\and         F.~                        Anders\inst{\ref{inst:0015}}
\and       R.I.~                      Anderson\orcit{0000-0001-8089-4419}\inst{\ref{inst:0150}}
\and         E.~                Anglada Varela\orcit{0000-0001-7563-0689}\inst{\ref{inst:0089}}
\and         T.~                        Antoja\orcit{0000-0003-2595-5148}\inst{\ref{inst:0015}}
\and         D.~                        Baines\orcit{0000-0002-6923-3756}\inst{\ref{inst:0063}}
\and       S.G.~                         Baker\orcit{0000-0002-6436-1257}\inst{\ref{inst:0057}}
\and         L.~        Balaguer-N\'{u}\~{n}ez\orcit{0000-0001-9789-7069}\inst{\ref{inst:0015}}
\and         E.~                      Balbinot\orcit{0000-0002-1322-3153}\inst{\ref{inst:0156}}
\and         Z.~                         Balog\orcit{0000-0003-1748-2926}\inst{\ref{inst:0008},\ref{inst:0029}}
\and         C.~                       Barache\inst{\ref{inst:0065}}
\and         D.~                       Barbato\inst{\ref{inst:0012},\ref{inst:0031}}
\and         M.~                        Barros\orcit{0000-0002-9728-9618}\inst{\ref{inst:0108}}
\and       M.A.~                       Barstow\orcit{0000-0002-7116-3259}\inst{\ref{inst:0163}}
\and         S.~                 Bartolom\'{e}\orcit{0000-0002-6290-6030}\inst{\ref{inst:0015}}
\and      J.-L.~                     Bassilana\inst{\ref{inst:0165}}
\and         N.~                       Bauchet\inst{\ref{inst:0005}}
\and         U.~                      Becciani\orcit{0000-0002-4389-8688}\inst{\ref{inst:0110}}
\and         M.~                    Bellazzini\orcit{0000-0001-8200-810X}\inst{\ref{inst:0038}}
\and         A.~                     Berihuete\orcit{0000-0002-8589-4423}\inst{\ref{inst:0169}}
\and         M.~                        Bernet\orcit{0000-0001-7503-1010}\inst{\ref{inst:0015}}
\and         S.~                       Bertone\orcit{0000-0001-9885-8440}\inst{\ref{inst:0171},\ref{inst:0172},\ref{inst:0031}}
\and         L.~                       Bianchi\orcit{0000-0002-7999-4372}\inst{\ref{inst:0174}}
\and         A.~                    Binnenfeld\orcit{0000-0002-9319-3838}\inst{\ref{inst:0175}}
\and         S.~               Blanco-Cuaresma\orcit{0000-0002-1584-0171}\inst{\ref{inst:0176}}
\and         A.~                       Blazere\inst{\ref{inst:0177}}
\and         T.~                          Boch\orcit{0000-0001-5818-2781}\inst{\ref{inst:0096}}
\and         A.~                       Bombrun\inst{\ref{inst:0179}}
\and         D.~                       Bossini\orcit{0000-0002-9480-8400}\inst{\ref{inst:0180}}
\and         S.~                    Bouquillon\inst{\ref{inst:0065},\ref{inst:0182}}
\and         A.~                     Bragaglia\orcit{0000-0002-0338-7883}\inst{\ref{inst:0038}}
\and         L.~                      Bramante\inst{\ref{inst:0048}}
\and         E.~                        Breedt\orcit{0000-0001-6180-3438}\inst{\ref{inst:0011}}
\and         A.~                       Bressan\orcit{0000-0002-7922-8440}\inst{\ref{inst:0186}}
\and         N.~                     Brouillet\orcit{0000-0002-3274-7024}\inst{\ref{inst:0010}}
\and         E.~                    Brugaletta\orcit{0000-0003-2598-6737}\inst{\ref{inst:0110}}
\and         B.~                   Bucciarelli\orcit{0000-0002-5303-0268}\inst{\ref{inst:0031},\ref{inst:0035}}
\and         A.~                       Burlacu\inst{\ref{inst:0191}}
\and       A.G.~                     Butkevich\orcit{0000-0002-4098-3588}\inst{\ref{inst:0031}}
\and         R.~                         Buzzi\orcit{0000-0001-9389-5701}\inst{\ref{inst:0031}}
\and         E.~                        Caffau\orcit{0000-0001-6011-6134}\inst{\ref{inst:0005}}
\and         R.~                   Cancelliere\orcit{0000-0002-9120-3799}\inst{\ref{inst:0195}}
\and         T.~                 Cantat-Gaudin\orcit{0000-0001-8726-2588}\inst{\ref{inst:0015},\ref{inst:0029}}
\and         R.~                      Carballo\orcit{0000-0001-7412-2498}\inst{\ref{inst:0198}}
\and         T.~                      Carlucci\inst{\ref{inst:0065}}
\and       M.I.~                     Carnerero\orcit{0000-0001-5843-5515}\inst{\ref{inst:0031}}
\and       J.M.~                      Carrasco\orcit{0000-0002-3029-5853}\inst{\ref{inst:0015}}
\and         L.~                   Casamiquela\orcit{0000-0001-5238-8674}\inst{\ref{inst:0010},\ref{inst:0005}}
\and         M.~                    Castellani\orcit{0000-0002-7650-7428}\inst{\ref{inst:0114}}
\and         A.~                 Castro-Ginard\orcit{0000-0002-9419-3725}\inst{\ref{inst:0002}}
\and         L.~                        Chaoul\inst{\ref{inst:0021}}
\and         P.~                       Charlot\orcit{0000-0002-9142-716X}\inst{\ref{inst:0010}}
\and         L.~                        Chemin\orcit{0000-0002-3834-7937}\inst{\ref{inst:0208}}
\and         V.~                    Chiaramida\inst{\ref{inst:0048}}
\and         A.~                     Chiavassa\orcit{0000-0003-3891-7554}\inst{\ref{inst:0009}}
\and         N.~                       Chornay\orcit{0000-0002-8767-3907}\inst{\ref{inst:0011}}
\and         G.~                     Comoretto\inst{\ref{inst:0063},\ref{inst:0213}}
\and         G.~                      Contursi\orcit{0000-0001-5370-1511}\inst{\ref{inst:0009}}
\and       W.J.~                        Cooper\orcit{0000-0003-3501-8967}\inst{\ref{inst:0215},\ref{inst:0031}}
\and         T.~                        Cornez\inst{\ref{inst:0165}}
\and         S.~                        Cowell\inst{\ref{inst:0011}}
\and         F.~                         Crifo\inst{\ref{inst:0005}}
\and         M.~                       Cropper\orcit{0000-0003-4571-9468}\inst{\ref{inst:0057}}
\and         M.~                        Crosta\orcit{0000-0003-4369-3786}\inst{\ref{inst:0031},\ref{inst:0222}}
\and         C.~                       Crowley\inst{\ref{inst:0179}}
\and         C.~                       Dafonte\orcit{0000-0003-4693-7555}\inst{\ref{inst:0091}}
\and         A.~                    Dapergolas\inst{\ref{inst:0069}}
\and         M.~                         David\orcit{0000-0002-4172-3112}\inst{\ref{inst:0226}}
\and         P.~                         David\inst{\ref{inst:0071}}
\and         P.~                    de Laverny\orcit{0000-0002-2817-4104}\inst{\ref{inst:0009}}
\and         F.~                      De Luise\orcit{0000-0002-6570-8208}\inst{\ref{inst:0229}}
\and         R.~                      De March\orcit{0000-0003-0567-842X}\inst{\ref{inst:0048}}
\and         J.~                     De Ridder\orcit{0000-0001-6726-2863}\inst{\ref{inst:0139}}
\and         R.~                      de Souza\inst{\ref{inst:0232}}
\and         A.~                     de Torres\inst{\ref{inst:0179}}
\and       E.F.~                    del Peloso\inst{\ref{inst:0008}}
\and         E.~                      del Pozo\inst{\ref{inst:0014}}
\and         M.~                         Delbo\orcit{0000-0002-8963-2404}\inst{\ref{inst:0009}}
\and         A.~                       Delgado\inst{\ref{inst:0088}}
\and      J.-B.~                       Delisle\orcit{0000-0001-5844-9888}\inst{\ref{inst:0012}}
\and         C.~                      Demouchy\inst{\ref{inst:0239}}
\and       T.E.~                 Dharmawardena\orcit{0000-0002-9583-5216}\inst{\ref{inst:0029}}
\and         P.~                     Di Matteo\inst{\ref{inst:0005}}
\and         S.~                       Diakite\inst{\ref{inst:0242}}
\and         C.~                        Diener\inst{\ref{inst:0011}}
\and         E.~                     Distefano\orcit{0000-0002-2448-2513}\inst{\ref{inst:0110}}
\and         C.~                       Dolding\inst{\ref{inst:0057}}
\and         B.~                    Edvardsson\inst{\ref{inst:0246}}
\and         H.~                          Enke\orcit{0000-0002-2366-8316}\inst{\ref{inst:0105}}
\and         C.~                         Fabre\inst{\ref{inst:0177}}
\and         M.~                      Fabrizio\orcit{0000-0001-5829-111X}\inst{\ref{inst:0114},\ref{inst:0115}}
\and         S.~                       Faigler\orcit{0000-0002-8368-5724}\inst{\ref{inst:0251}}
\and         G.~                      Fedorets\orcit{0000-0002-8418-4809}\inst{\ref{inst:0117},\ref{inst:0253}}
\and         P.~                      Fernique\orcit{0000-0002-3304-2923}\inst{\ref{inst:0096},\ref{inst:0255}}
\and         A.~                        Fienga\orcit{0000-0002-4755-7637}\inst{\ref{inst:0256},\ref{inst:0071}}
\and         F.~                      Figueras\orcit{0000-0002-3393-0007}\inst{\ref{inst:0015}}
\and         Y.~                      Fournier\orcit{0000-0002-6633-9088}\inst{\ref{inst:0105}}
\and         C.~                        Fouron\inst{\ref{inst:0191}}
\and         F.~                     Fragkoudi\orcit{0000-0002-0897-3013}\inst{\ref{inst:0261},\ref{inst:0262},\ref{inst:0263}}
\and         M.~                           Gai\orcit{0000-0001-9008-134X}\inst{\ref{inst:0031}}
\and         A.~              Garcia-Gutierrez\inst{\ref{inst:0015}}
\and         M.~              Garcia-Reinaldos\inst{\ref{inst:0013}}
\and         M.~             Garc\'{i}a-Torres\orcit{0000-0002-6867-7080}\inst{\ref{inst:0267}}
\and         A.~                      Garofalo\orcit{0000-0002-5907-0375}\inst{\ref{inst:0038}}
\and         A.~                         Gavel\orcit{0000-0002-2963-722X}\inst{\ref{inst:0046}}
\and         P.~                        Gavras\orcit{0000-0002-4383-4836}\inst{\ref{inst:0088}}
\and         E.~                       Gerlach\orcit{0000-0002-9533-2168}\inst{\ref{inst:0016}}
\and         R.~                         Geyer\orcit{0000-0001-6967-8707}\inst{\ref{inst:0016}}
\and         P.~                      Giacobbe\orcit{0000-0001-7034-7024}\inst{\ref{inst:0031}}
\and         G.~                       Gilmore\orcit{0000-0003-4632-0213}\inst{\ref{inst:0011}}
\and         S.~                        Girona\orcit{0000-0002-1975-1918}\inst{\ref{inst:0275}}
\and         G.~                     Giuffrida\inst{\ref{inst:0114}}
\and         R.~                         Gomel\inst{\ref{inst:0251}}
\and         A.~                         Gomez\orcit{0000-0002-3796-3690}\inst{\ref{inst:0091}}
\and         J.~    Gonz\'{a}lez-N\'{u}\~{n}ez\orcit{0000-0001-5311-5555}\inst{\ref{inst:0079},\ref{inst:0280}}
\and         I.~   Gonz\'{a}lez-Santamar\'{i}a\orcit{0000-0002-8537-9384}\inst{\ref{inst:0091}}
\and       J.J.~            Gonz\'{a}lez-Vidal\inst{\ref{inst:0015}}
\and         M.~                       Granvik\orcit{0000-0002-5624-1888}\inst{\ref{inst:0117},\ref{inst:0284}}
\and         P.~                      Guillout\inst{\ref{inst:0096}}
\and         J.~                       Guiraud\inst{\ref{inst:0021}}
\and         R.~     Guti\'{e}rrez-S\'{a}nchez\inst{\ref{inst:0063}}
\and       L.P.~                           Guy\orcit{0000-0003-0800-8755}\inst{\ref{inst:0052},\ref{inst:0289}}
\and         D.~                Hatzidimitriou\orcit{0000-0002-5415-0464}\inst{\ref{inst:0290},\ref{inst:0069}}
\and         M.~                        Hauser\inst{\ref{inst:0029},\ref{inst:0293}}
\and         M.~                       Haywood\orcit{0000-0003-0434-0400}\inst{\ref{inst:0005}}
\and         A.~                        Helmer\inst{\ref{inst:0165}}
\and         A.~                         Helmi\orcit{0000-0003-3937-7641}\inst{\ref{inst:0156}}
\and       M.H.~                     Sarmiento\orcit{0000-0003-4252-5115}\inst{\ref{inst:0014}}
\and       S.L.~                       Hidalgo\orcit{0000-0002-0002-9298}\inst{\ref{inst:0298},\ref{inst:0299}}
\and       T.~                          Hilger\orcit{0000-0003-1646-0063}\inst{\ref{inst:0016}}
\and         N.~                   H\l{}adczuk\orcit{0000-0001-9163-4209}\inst{\ref{inst:0013},\ref{inst:0301}}
\and         D.~                         Hobbs\orcit{0000-0002-2696-1366}\inst{\ref{inst:0018}}
\and         G.~                       Holland\inst{\ref{inst:0011}}
\and       H.E.~                        Huckle\inst{\ref{inst:0057}}
\and         K.~                       Jardine\inst{\ref{inst:0305}}
\and         G.~                    Jasniewicz\inst{\ref{inst:0306}}
\and         A.~          Jean-Antoine Piccolo\orcit{0000-0001-8622-212X}\inst{\ref{inst:0021}}
\and     \'{O}.~            Jim\'{e}nez-Arranz\orcit{0000-0001-7434-5165}\inst{\ref{inst:0015}}
\and         A.~                      Jorissen\orcit{0000-0002-1883-4578}\inst{\ref{inst:0022}}
\and         J.~             Juaristi Campillo\inst{\ref{inst:0008}}
\and         F.~                         Julbe\inst{\ref{inst:0015}}
\and         L.~                     Karbevska\inst{\ref{inst:0052},\ref{inst:0313}}
\and         P.~                      Kervella\orcit{0000-0003-0626-1749}\inst{\ref{inst:0314}}
\and         S.~                        Khanna\orcit{0000-0002-2604-4277}\inst{\ref{inst:0156},\ref{inst:0031}}
\and         M.~                      Kontizas\orcit{0000-0001-7177-0158}\inst{\ref{inst:0290}}
\and         G.~                    Kordopatis\orcit{0000-0002-9035-3920}\inst{\ref{inst:0009}}
\and       A.J.~                          Korn\orcit{0000-0002-3881-6756}\inst{\ref{inst:0046}}
\and      \'{A}~                K\'{o}sp\'{a}l\orcit{'{u}t 15-17, 1121 B}\inst{\ref{inst:0136},\ref{inst:0029},\ref{inst:0137}}
\and         Z.~           Kostrzewa-Rutkowska\inst{\ref{inst:0002},\ref{inst:0324}}
\and         K.~                Kruszy\'{n}ska\orcit{0000-0002-2729-5369}\inst{\ref{inst:0325}}
\and         M.~                           Kun\orcit{0000-0002-7538-5166}\inst{\ref{inst:0136}}
\and         P.~                       Laizeau\inst{\ref{inst:0327}}
\and         S.~                       Lambert\orcit{0000-0001-6759-5502}\inst{\ref{inst:0065}}
\and       A.F.~                         Lanza\orcit{0000-0001-5928-7251}\inst{\ref{inst:0110}}
\and         Y.~                         Lasne\inst{\ref{inst:0165}}
\and      J.-F.~                    Le Campion\inst{\ref{inst:0010}}
\and         Y.~                      Lebreton\orcit{0000-0002-4834-2144}\inst{\ref{inst:0314},\ref{inst:0333}}
\and         T.~                     Lebzelter\orcit{0000-0002-0702-7551}\inst{\ref{inst:0148}}
\and         S.~                        Leccia\orcit{0000-0001-5685-6930}\inst{\ref{inst:0335}}
\and         N.~                       Leclerc\inst{\ref{inst:0005}}
\and         I.~                 Lecoeur-Taibi\orcit{0000-0003-0029-8575}\inst{\ref{inst:0052}}
\and         S.~                          Liao\orcit{0000-0002-9346-0211}\inst{\ref{inst:0338},\ref{inst:0031},\ref{inst:0340}}
\and       E.L.~                        Licata\orcit{0000-0002-5203-0135}\inst{\ref{inst:0031}}
\and     H.E.P.~                  Lindstr{\o}m\inst{\ref{inst:0031},\ref{inst:0343},\ref{inst:0344}}
\and       T.A.~                        Lister\orcit{0000-0002-3818-7769}\inst{\ref{inst:0345}}
\and         E.~                       Livanou\orcit{0000-0003-0628-2347}\inst{\ref{inst:0290}}
\and         A.~                         Lobel\orcit{0000-0001-5030-019X}\inst{\ref{inst:0043}}
\and         A.~                         Lorca\inst{\ref{inst:0014}}
\and         C.~                          Loup\inst{\ref{inst:0096}}
\and         P.~                 Madrero Pardo\inst{\ref{inst:0015}}
\and         A.~               Magdaleno Romeo\inst{\ref{inst:0191}}
\and         S.~                       Managau\inst{\ref{inst:0165}}
\and       R.G.~                          Mann\orcit{0000-0002-0194-325X}\inst{\ref{inst:0083}}
\and         M.~                      Manteiga\orcit{0000-0002-7711-5581}\inst{\ref{inst:0354}}
\and       J.M.~                      Marchant\orcit{0000-0002-3678-3145}\inst{\ref{inst:0355}}
\and         M.~                       Marconi\orcit{0000-0002-1330-2927}\inst{\ref{inst:0335}}
\and         J.~                        Marcos\inst{\ref{inst:0063}}
\and     M.M.S.~                 Marcos Santos\inst{\ref{inst:0079}}
\and         D.~                Mar\'{i}n Pina\orcit{0000-0001-6482-1842}\inst{\ref{inst:0015}}
\and         S.~                      Marinoni\orcit{0000-0001-7990-6849}\inst{\ref{inst:0114},\ref{inst:0115}}
\and         F.~                       Marocco\orcit{0000-0001-7519-1700}\inst{\ref{inst:0362}}
\and       D.J.~                      Marshall\orcit{0000-0003-3956-3524}\inst{\ref{inst:0363}}
\and         L.~                   Martin Polo\inst{\ref{inst:0079}}
\and       J.M.~            Mart\'{i}n-Fleitas\orcit{0000-0002-8594-569X}\inst{\ref{inst:0014}}
\and         G.~                        Marton\orcit{0000-0002-1326-1686}\inst{\ref{inst:0136}}
\and         N.~                          Mary\inst{\ref{inst:0165}}
\and         A.~                         Masip\orcit{0000-0003-1419-0020}\inst{\ref{inst:0015}}
\and         D.~                       Massari\orcit{0000-0001-8892-4301}\inst{\ref{inst:0038}}
\and         A.~          Mastrobuono-Battisti\orcit{0000-0002-2386-9142}\inst{\ref{inst:0005}}
\and         T.~                         Mazeh\orcit{0000-0002-3569-3391}\inst{\ref{inst:0251}}
\and       P.J.~                      McMillan\orcit{0000-0002-8861-2620}\inst{\ref{inst:0018}}
\and         S.~                       Messina\orcit{0000-0002-2851-2468}\inst{\ref{inst:0110}}
\and         D.~                      Michalik\orcit{0000-0002-7618-6556}\inst{\ref{inst:0003}}
\and       N.R.~                        Millar\inst{\ref{inst:0011}}
\and         A.~                         Mints\orcit{0000-0002-8440-1455}\inst{\ref{inst:0105}}
\and         D.~                        Molina\orcit{0000-0003-4814-0275}\inst{\ref{inst:0015}}
\and         R.~                      Molinaro\orcit{0000-0003-3055-6002}\inst{\ref{inst:0335}}
\and         L.~                    Moln\'{a}r\orcit{0000-0002-8159-1599}\inst{\ref{inst:0136},\ref{inst:0380},\ref{inst:0137}}
\and         G.~                        Monari\orcit{0000-0002-6863-0661}\inst{\ref{inst:0096}}
\and         M.~                   Mongui\'{o}\orcit{0000-0002-4519-6700}\inst{\ref{inst:0015}}
\and         P.~                   Montegriffo\orcit{0000-0001-5013-5948}\inst{\ref{inst:0038}}
\and         A.~                       Montero\inst{\ref{inst:0014}}
\and         R.~                           Mor\orcit{0000-0002-8179-6527}\inst{\ref{inst:0015}}
\and         A.~                          Mora\inst{\ref{inst:0014}}
\and         R.~                    Morbidelli\orcit{0000-0001-7627-4946}\inst{\ref{inst:0031}}
\and         T.~                         Morel\orcit{0000-0002-8176-4816}\inst{\ref{inst:0081}}
\and         D.~                        Morris\inst{\ref{inst:0083}}
\and         T.~                      Muraveva\orcit{0000-0002-0969-1915}\inst{\ref{inst:0038}}
\and       C.P.~                        Murphy\inst{\ref{inst:0013}}
\and         I.~                       Musella\orcit{0000-0001-5909-6615}\inst{\ref{inst:0335}}
\and         Z.~                          Nagy\orcit{0000-0002-3632-1194}\inst{\ref{inst:0136}}
\and         L.~                         Noval\inst{\ref{inst:0165}}
\and         F.~                     Oca\~{n}a\inst{\ref{inst:0063},\ref{inst:0396}}
\and         A.~                         Ogden\inst{\ref{inst:0011}}
\and         C.~                     Ordenovic\inst{\ref{inst:0009}}
\and       J.O.~                        Osinde\inst{\ref{inst:0088}}
\and         C.~                        Pagani\orcit{0000-0001-5477-4720}\inst{\ref{inst:0163}}
\and         I.~                        Pagano\orcit{0000-0001-9573-4928}\inst{\ref{inst:0110}}
\and         L.~                     Palaversa\orcit{0000-0003-3710-0331}\inst{\ref{inst:0402},\ref{inst:0011}}
\and       P.A.~                       Palicio\orcit{0000-0002-7432-8709}\inst{\ref{inst:0009}}
\and         L.~               Pallas-Quintela\orcit{0000-0001-9296-3100}\inst{\ref{inst:0091}}
\and         A.~                        Panahi\orcit{0000-0001-5850-4373}\inst{\ref{inst:0251}}
\and         S.~               Payne-Wardenaar\inst{\ref{inst:0008}}
\and         X.~         Pe\~{n}alosa Esteller\inst{\ref{inst:0015}}
\and         A.~                 Penttil\"{ a}\orcit{0000-0001-7403-1721}\inst{\ref{inst:0117}}
\and         B.~                        Pichon\orcit{0000 0000 0062 1449}\inst{\ref{inst:0009}}
\and       A.M.~                    Piersimoni\orcit{0000-0002-8019-3708}\inst{\ref{inst:0229}}
\and      F.-X.~                        Pineau\orcit{0000-0002-2335-4499}\inst{\ref{inst:0096}}
\and         E.~                        Plachy\orcit{0000-0002-5481-3352}\inst{\ref{inst:0136},\ref{inst:0380},\ref{inst:0137}}
\and         G.~                          Plum\inst{\ref{inst:0005}}
\and         E.~                        Poggio\orcit{0000-0003-3793-8505}\inst{\ref{inst:0009},\ref{inst:0031}}
\and         A.~                      Pr\v{s}a\orcit{0000-0002-1913-0281}\inst{\ref{inst:0418}}
\and         L.~                        Pulone\orcit{0000-0002-5285-998X}\inst{\ref{inst:0114}}
\and         E.~                        Racero\orcit{0000-0002-6101-9050}\inst{\ref{inst:0079},\ref{inst:0396}}
\and         S.~                       Ragaini\inst{\ref{inst:0038}}
\and         M.~                        Rainer\orcit{0000-0002-8786-2572}\inst{\ref{inst:0024},\ref{inst:0424}}
\and       C.M.~                       Raiteri\orcit{0000-0003-1784-2784}\inst{\ref{inst:0031}}
\and         N.~                       Rambaux\orcit{0000-0002-9380-271X}\inst{\ref{inst:0071}}
\and         P.~                         Ramos\orcit{0000-0002-5080-7027}\inst{\ref{inst:0015},\ref{inst:0096}}
\and         M.~                  Ramos-Lerate\inst{\ref{inst:0063}}
\and         P.~                  Re Fiorentin\orcit{0000-0002-4995-0475}\inst{\ref{inst:0031}}
\and         S.~                        Regibo\inst{\ref{inst:0139}}
\and       P.J.~                      Richards\inst{\ref{inst:0432}}
\and         C.~                     Rios Diaz\inst{\ref{inst:0088}}
\and         V.~                        Ripepi\orcit{0000-0003-1801-426X}\inst{\ref{inst:0335}}
\and         A.~                          Riva\orcit{0000-0002-6928-8589}\inst{\ref{inst:0031}}
\and      H.-W.~                           Rix\orcit{0000-0003-4996-9069}\inst{\ref{inst:0029}}
\and         G.~                         Rixon\orcit{0000-0003-4399-6568}\inst{\ref{inst:0011}}
\and         N.~                      Robichon\orcit{0000-0003-4545-7517}\inst{\ref{inst:0005}}
\and       A.C.~                         Robin\orcit{0000-0001-8654-9499}\inst{\ref{inst:0124}}
\and         C.~                         Robin\inst{\ref{inst:0165}}
\and         M.~                       Roelens\orcit{0000-0003-0876-4673}\inst{\ref{inst:0012}}
\and     H.R.O.~                        Rogues\inst{\ref{inst:0239}}
\and         L.~                    Rohrbasser\inst{\ref{inst:0052}}
\and         M.~              Romero-G\'{o}mez\orcit{0000-0003-3936-1025}\inst{\ref{inst:0015}}
\and         N.~                        Rowell\orcit{0000-0003-3809-1895}\inst{\ref{inst:0083}}
\and         F.~                         Royer\orcit{0000-0002-9374-8645}\inst{\ref{inst:0005}}
\and         D.~                    Ruz Mieres\orcit{0000-0002-9455-157X}\inst{\ref{inst:0011}}
\and       K.A.~                       Rybicki\orcit{0000-0002-9326-9329}\inst{\ref{inst:0325}}
\and         G.~                      Sadowski\orcit{0000-0002-3411-1003}\inst{\ref{inst:0022}}
\and         A.~        S\'{a}ez N\'{u}\~{n}ez\inst{\ref{inst:0015}}
\and         A.~       Sagrist\`{a} Sell\'{e}s\orcit{0000-0001-6191-2028}\inst{\ref{inst:0008}}
\and         J.~                      Sahlmann\orcit{0000-0001-9525-3673}\inst{\ref{inst:0088}}
\and         E.~                      Salguero\inst{\ref{inst:0089}}
\and         N.~                       Samaras\orcit{0000-0001-8375-6652}\inst{\ref{inst:0043},\ref{inst:0455}}
\and         V.~               Sanchez Gimenez\orcit{0000-0003-1797-3557}\inst{\ref{inst:0015}}
\and         N.~                         Sanna\orcit{0000-0001-9275-9492}\inst{\ref{inst:0024}}
\and         R.~                 Santove\~{n}a\orcit{0000-0002-9257-2131}\inst{\ref{inst:0091}}
\and         M.~                       Sarasso\orcit{0000-0001-5121-0727}\inst{\ref{inst:0031}}
\and        M.~                    Schultheis\orcit{0000-0002-6590-1657}\inst{\ref{inst:0009}}
\and         E.~                       Sciacca\orcit{0000-0002-5574-2787}\inst{\ref{inst:0110}}
\and         M.~                         Segol\inst{\ref{inst:0239}}
\and       J.C.~                       Segovia\inst{\ref{inst:0079}}
\and         D.~                 S\'{e}gransan\orcit{0000-0003-2355-8034}\inst{\ref{inst:0012}}
\and         D.~                        Semeux\inst{\ref{inst:0177}}
\and         S.~                        Shahaf\orcit{0000-0001-9298-8068}\inst{\ref{inst:0466}}
\and       H.I.~                      Siddiqui\orcit{0000-0003-1853-6033}\inst{\ref{inst:0467}}
\and         A.~                       Siebert\orcit{0000-0001-8059-2840}\inst{\ref{inst:0096},\ref{inst:0255}}
\and         L.~                       Siltala\orcit{0000-0002-6938-794X}\inst{\ref{inst:0117}}
\and         A.~                       Silvelo\orcit{0000-0002-5126-6365}\inst{\ref{inst:0091}}
\and         E.~                        Slezak\inst{\ref{inst:0009}}
\and         I.~                        Slezak\inst{\ref{inst:0009}}
\and       R.L.~                         Smart\orcit{0000-0002-4424-4766}\inst{\ref{inst:0031}}
\and       O.N.~                        Snaith\inst{\ref{inst:0005}}
\and         E.~                        Solano\inst{\ref{inst:0476}}
\and         F.~                       Solitro\inst{\ref{inst:0048}}
\and         D.~                        Souami\orcit{0000-0003-4058-0815}\inst{\ref{inst:0314},\ref{inst:0479}}
\and         J.~                       Souchay\inst{\ref{inst:0065}}
\and         A.~                        Spagna\orcit{0000-0003-1732-2412}\inst{\ref{inst:0031}}
\and         L.~                         Spina\orcit{0000-0002-9760-6249}\inst{\ref{inst:0001}}
\and         F.~                         Spoto\orcit{0000-0001-7319-5847}\inst{\ref{inst:0176}}
\and       I.A.~                        Steele\orcit{0000-0001-8397-5759}\inst{\ref{inst:0355}}
\and         H.~            Steidelm\"{ u}ller\inst{\ref{inst:0016}}
\and       C.A.~                    Stephenson\inst{\ref{inst:0063},\ref{inst:0487}}
\and         M.~                  S\"{ u}veges\orcit{0000-0003-3017-5322}\inst{\ref{inst:0488}}
\and         J.~                        Surdej\orcit{0000-0002-7005-1976}\inst{\ref{inst:0081},\ref{inst:0490}}
\and         L.~                      Szabados\orcit{0000-0002-2046-4131}\inst{\ref{inst:0136}}
\and         E.~                  Szegedi-Elek\orcit{0000-0001-7807-6644}\inst{\ref{inst:0136}}
\and         F.~                         Taris\inst{\ref{inst:0065}}
\and       M.B.~                        Taylor\orcit{0000-0002-4209-1479}\inst{\ref{inst:0494}}
\and         R.~                      Teixeira\orcit{0000-0002-6806-6626}\inst{\ref{inst:0232}}
\and         L.~                       Tolomei\orcit{0000-0002-3541-3230}\inst{\ref{inst:0048}}
\and         N.~                       Tonello\orcit{0000-0003-0550-1667}\inst{\ref{inst:0275}}
\and         F.~                         Torra\orcit{0000-0002-8429-299X}\inst{\ref{inst:0039}}
\and         J.~               Torra$^\dagger$\inst{\ref{inst:0015}}
\and         G.~                Torralba Elipe\orcit{0000-0001-8738-194X}\inst{\ref{inst:0091}}
\and         M.~                     Trabucchi\orcit{0000-0002-1429-2388}\inst{\ref{inst:0501},\ref{inst:0012}}
\and       A.T.~                       Tsounis\inst{\ref{inst:0503}}
\and         C.~                         Turon\orcit{0000-0003-1236-5157}\inst{\ref{inst:0005}}
\and         A.~                          Ulla\orcit{0000-0001-6424-5005}\inst{\ref{inst:0505}}
\and         N.~                         Unger\orcit{0000-0003-3993-7127}\inst{\ref{inst:0012}}
\and       M.V.~                      Vaillant\inst{\ref{inst:0165}}
\and         E.~                    van Dillen\inst{\ref{inst:0239}}
\and         W.~                    van Reeven\inst{\ref{inst:0509}}
\and         O.~                         Vanel\orcit{0000-0002-7898-0454}\inst{\ref{inst:0005}}
\and         A.~                     Vecchiato\orcit{0000-0003-1399-5556}\inst{\ref{inst:0031}}
\and         Y.~                         Viala\inst{\ref{inst:0005}}
\and         D.~                       Vicente\orcit{0000-0002-1584-1182}\inst{\ref{inst:0275}}
\and         S.~                     Voutsinas\inst{\ref{inst:0083}}
\and         M.~                        Weiler\inst{\ref{inst:0015}}
\and         T.~                        Wevers\orcit{0000-0002-4043-9400}\inst{\ref{inst:0011},\ref{inst:0517}}
\and      \L{}.~                   Wyrzykowski\orcit{0000-0002-9658-6151}\inst{\ref{inst:0325}}
\and         A.~                        Yoldas\inst{\ref{inst:0011}}
\and         P.~                         Yvard\inst{\ref{inst:0239}}
\and         H.~                          Zhao\orcit{0000-0003-2645-6869}\inst{\ref{inst:0009}}
\and         J.~                         Zorec\inst{\ref{inst:0522}}
\and         S.~                        Zucker\orcit{0000-0003-3173-3138}\inst{\ref{inst:0175}}
\and         T.~                       Zwitter\orcit{0000-0002-2325-8763}\inst{\ref{inst:0524}}
}
\institute{
     INAF - Osservatorio astronomico di Padova, Vicolo Osservatorio 5, 35122 Padova, Italy\relax                                                                                                                                                                                                                                                                   \label{inst:0001}
\and Leiden Observatory, Leiden University, Niels Bohrweg 2, 2333 CA Leiden, The Netherlands\relax                                                                                                                                                                                                                                                                 \label{inst:0002}\vfill
\and European Space Agency (ESA), European Space Research and Technology Centre (ESTEC), Keplerlaan 1, 2201AZ, Noordwijk, The Netherlands\relax                                                                                                                                                                                                                    \label{inst:0003}\vfill
\and GEPI, Observatoire de Paris, Universit\'{e} PSL, CNRS, 5 Place Jules Janssen, 92190 Meudon, France\relax                                                                                                                                                                                                                                                      \label{inst:0005}\vfill
\and Univ. Grenoble Alpes, CNRS, IPAG, 38000 Grenoble, France\relax                                                                                                                                                                                                                                                                                                \label{inst:0006}\vfill
\and Astronomisches Rechen-Institut, Zentrum f\"{ u}r Astronomie der Universit\"{ a}t Heidelberg, M\"{ o}nchhofstr. 12-14, 69120 Heidelberg, Germany\relax                                                                                                                                                                                                         \label{inst:0008}\vfill
\and Universit\'{e} C\^{o}te d'Azur, Observatoire de la C\^{o}te d'Azur, CNRS, Laboratoire Lagrange, Bd de l'Observatoire, CS 34229, 06304 Nice Cedex 4, France\relax                                                                                                                                                                                              \label{inst:0009}\vfill
\and Laboratoire d'astrophysique de Bordeaux, Univ. Bordeaux, CNRS, B18N, all{\'e}e Geoffroy Saint-Hilaire, 33615 Pessac, France\relax                                                                                                                                                                                                                             \label{inst:0010}\vfill
\and Institute of Astronomy, University of Cambridge, Madingley Road, Cambridge CB3 0HA, United Kingdom\relax                                                                                                                                                                                                                                                      \label{inst:0011}\vfill
\and Department of Astronomy, University of Geneva, Chemin Pegasi 51, 1290 Versoix, Switzerland\relax                                                                                                                                                                                                                                                              \label{inst:0012}\vfill
\and European Space Agency (ESA), European Space Astronomy Centre (ESAC), Camino bajo del Castillo, s/n, Urbanizacion Villafranca del Castillo, Villanueva de la Ca\~{n}ada, 28692 Madrid, Spain\relax                                                                                                                                                             \label{inst:0013}\vfill
\and Aurora Technology for European Space Agency (ESA), Camino bajo del Castillo, s/n, Urbanizacion Villafranca del Castillo, Villanueva de la Ca\~{n}ada, 28692 Madrid, Spain\relax                                                                                                                                                                               \label{inst:0014}\vfill
\and Institut de Ci\`{e}ncies del Cosmos (ICCUB), Universitat  de  Barcelona  (IEEC-UB), Mart\'{i} i  Franqu\`{e}s  1, 08028 Barcelona, Spain\relax                                                                                                                                                                                                                \label{inst:0015}\vfill
\and Lohrmann Observatory, Technische Universit\"{ a}t Dresden, Mommsenstra{\ss}e 13, 01062 Dresden, Germany\relax                                                                                                                                                                                                                                                 \label{inst:0016}\vfill
\and Lund Observatory, Department of Astronomy and Theoretical Physics, Lund University, Box 43, 22100 Lund, Sweden\relax                                                                                                                                                                                                                                          \label{inst:0018}\vfill
\and CNES Centre Spatial de Toulouse, 18 avenue Edouard Belin, 31401 Toulouse Cedex 9, France\relax                                                                                                                                                                                                                                                                \label{inst:0021}\vfill
\and Institut d'Astronomie et d'Astrophysique, Universit\'{e} Libre de Bruxelles CP 226, Boulevard du Triomphe, 1050 Brussels, Belgium\relax                                                                                                                                                                                                                       \label{inst:0022}\vfill
\and F.R.S.-FNRS, Rue d'Egmont 5, 1000 Brussels, Belgium\relax                                                                                                                                                                                                                                                                                                     \label{inst:0023}\vfill
\and INAF - Osservatorio Astrofisico di Arcetri, Largo Enrico Fermi 5, 50125 Firenze, Italy\relax                                                                                                                                                                                                                                                                  \label{inst:0024}\vfill
\and Max Planck Institute for Astronomy, K\"{ o}nigstuhl 17, 69117 Heidelberg, Germany\relax                                                                                                                                                                                                                                                                       \label{inst:0029}\vfill
\and INAF - Osservatorio Astrofisico di Torino, via Osservatorio 20, 10025 Pino Torinese (TO), Italy\relax                                                                                                                                                                                                                                                         \label{inst:0031}\vfill
\and European Space Agency (ESA, retired)\relax                                                                                                                                                                                                                                                                                                                    \label{inst:0032}\vfill
\and University of Turin, Department of Physics, Via Pietro Giuria 1, 10125 Torino, Italy\relax                                                                                                                                                                                                                                                                    \label{inst:0035}\vfill
\and INAF - Osservatorio di Astrofisica e Scienza dello Spazio di Bologna, via Piero Gobetti 93/3, 40129 Bologna, Italy\relax                                                                                                                                                                                                                                      \label{inst:0038}\vfill
\and DAPCOM for Institut de Ci\`{e}ncies del Cosmos (ICCUB), Universitat  de  Barcelona  (IEEC-UB), Mart\'{i} i  Franqu\`{e}s  1, 08028 Barcelona, Spain\relax                                                                                                                                                                                                     \label{inst:0039}\vfill
\and Royal Observatory of Belgium, Ringlaan 3, 1180 Brussels, Belgium\relax                                                                                                                                                                                                                                                                                        \label{inst:0043}\vfill
\and Observational Astrophysics, Division of Astronomy and Space Physics, Department of Physics and Astronomy, Uppsala University, Box 516, 751 20 Uppsala, Sweden\relax                                                                                                                                                                                           \label{inst:0046}\vfill
\and ALTEC S.p.a, Corso Marche, 79,10146 Torino, Italy\relax                                                                                                                                                                                                                                                                                                       \label{inst:0048}\vfill
\and S\`{a}rl, Geneva, Switzerland\relax                                                                                                                                                                                                                                                                                                                           \label{inst:0051}\vfill
\and Department of Astronomy, University of Geneva, Chemin d'Ecogia 16, 1290 Versoix, Switzerland\relax                                                                                                                                                                                                                                                            \label{inst:0052}\vfill
\and Mullard Space Science Laboratory, University College London, Holmbury St Mary, Dorking, Surrey RH5 6NT, United Kingdom\relax                                                                                                                                                                                                                                  \label{inst:0057}\vfill
\and Gaia DPAC Project Office, ESAC, Camino bajo del Castillo, s/n, Urbanizacion Villafranca del Castillo, Villanueva de la Ca\~{n}ada, 28692 Madrid, Spain\relax                                                                                                                                                                                                  \label{inst:0060}\vfill
\and Telespazio UK S.L. for European Space Agency (ESA), Camino bajo del Castillo, s/n, Urbanizacion Villafranca del Castillo, Villanueva de la Ca\~{n}ada, 28692 Madrid, Spain\relax                                                                                                                                                                              \label{inst:0063}\vfill
\and SYRTE, Observatoire de Paris, Universit\'{e} PSL, CNRS,  Sorbonne Universit\'{e}, LNE, 61 avenue de l'Observatoire 75014 Paris, France\relax                                                                                                                                                                                                                  \label{inst:0065}\vfill
\and National Observatory of Athens, I. Metaxa and Vas. Pavlou, Palaia Penteli, 15236 Athens, Greece\relax                                                                                                                                                                                                                                                         \label{inst:0069}\vfill
\and IMCCE, Observatoire de Paris, Universit\'{e} PSL, CNRS, Sorbonne Universit{\'e}, Univ. Lille, 77 av. Denfert-Rochereau, 75014 Paris, France\relax                                                                                                                                                                                                             \label{inst:0071}\vfill
\and Serco Gesti\'{o}n de Negocios for European Space Agency (ESA), Camino bajo del Castillo, s/n, Urbanizacion Villafranca del Castillo, Villanueva de la Ca\~{n}ada, 28692 Madrid, Spain\relax                                                                                                                                                                   \label{inst:0079}\vfill
\and Institut d'Astrophysique et de G\'{e}ophysique, Universit\'{e} de Li\`{e}ge, 19c, All\'{e}e du 6 Ao\^{u}t, B-4000 Li\`{e}ge, Belgium\relax                                                                                                                                                                                                                    \label{inst:0081}\vfill
\and CRAAG - Centre de Recherche en Astronomie, Astrophysique et G\'{e}ophysique, Route de l'Observatoire Bp 63 Bouzareah 16340 Algiers, Algeria\relax                                                                                                                                                                                                             \label{inst:0082}\vfill
\and Institute for Astronomy, University of Edinburgh, Royal Observatory, Blackford Hill, Edinburgh EH9 3HJ, United Kingdom\relax                                                                                                                                                                                                                                  \label{inst:0083}\vfill
\and RHEA for European Space Agency (ESA), Camino bajo del Castillo, s/n, Urbanizacion Villafranca del Castillo, Villanueva de la Ca\~{n}ada, 28692 Madrid, Spain\relax                                                                                                                                                                                            \label{inst:0088}\vfill
\and ATG Europe for European Space Agency (ESA), Camino bajo del Castillo, s/n, Urbanizacion Villafranca del Castillo, Villanueva de la Ca\~{n}ada, 28692 Madrid, Spain\relax                                                                                                                                                                                      \label{inst:0089}\vfill
\and CIGUS CITIC - Department of Computer Science and Information Technologies, University of A Coru\~{n}a, Campus de Elvi\~{n}a s/n, A Coru\~{n}a, 15071, Spain\relax                                                                                                                                                                                             \label{inst:0091}\vfill
\and Universit\'{e} de Strasbourg, CNRS, Observatoire astronomique de Strasbourg, UMR 7550,  11 rue de l'Universit\'{e}, 67000 Strasbourg, France\relax                                                                                                                                                                                                            \label{inst:0096}\vfill
\and Kavli Institute for Cosmology Cambridge, Institute of Astronomy, Madingley Road, Cambridge, CB3 0HA\relax                                                                                                                                                                                                                                                     \label{inst:0099}\vfill
\and Leibniz Institute for Astrophysics Potsdam (AIP), An der Sternwarte 16, 14482 Potsdam, Germany\relax                                                                                                                                                                                                                                                          \label{inst:0105}\vfill
\and CENTRA, Faculdade de Ci\^{e}ncias, Universidade de Lisboa, Edif. C8, Campo Grande, 1749-016 Lisboa, Portugal\relax                                                                                                                                                                                                                                            \label{inst:0108}\vfill
\and Department of Informatics, Donald Bren School of Information and Computer Sciences, University of California, Irvine, 5226 Donald Bren Hall, 92697-3440 CA Irvine, United States\relax                                                                                                                                                                        \label{inst:0109}\vfill
\and INAF - Osservatorio Astrofisico di Catania, via S. Sofia 78, 95123 Catania, Italy\relax                                                                                                                                                                                                                                                                       \label{inst:0110}\vfill
\and Dipartimento di Fisica e Astronomia ""Ettore Majorana"", Universit\`{a} di Catania, Via S. Sofia 64, 95123 Catania, Italy\relax                                                                                                                                                                                                                               \label{inst:0111}\vfill
\and INAF - Osservatorio Astronomico di Roma, Via Frascati 33, 00078 Monte Porzio Catone (Roma), Italy\relax                                                                                                                                                                                                                                                       \label{inst:0114}\vfill
\and Space Science Data Center - ASI, Via del Politecnico SNC, 00133 Roma, Italy\relax                                                                                                                                                                                                                                                                             \label{inst:0115}\vfill
\and Department of Physics, University of Helsinki, P.O. Box 64, 00014 Helsinki, Finland\relax                                                                                                                                                                                                                                                                     \label{inst:0117}\vfill
\and Finnish Geospatial Research Institute FGI, Geodeetinrinne 2, 02430 Masala, Finland\relax                                                                                                                                                                                                                                                                      \label{inst:0118}\vfill
\and Institut UTINAM CNRS UMR6213, Universit\'{e} Bourgogne Franche-Comt\'{e}, OSU THETA Franche-Comt\'{e} Bourgogne, Observatoire de Besan\c{c}on, BP1615, 25010 Besan\c{c}on Cedex, France\relax                                                                                                                                                                 \label{inst:0124}\vfill
\and HE Space Operations BV for European Space Agency (ESA), Keplerlaan 1, 2201AZ, Noordwijk, The Netherlands\relax                                                                                                                                                                                                                                                \label{inst:0127}\vfill
\and Dpto. de Inteligencia Artificial, UNED, c/ Juan del Rosal 16, 28040 Madrid, Spain\relax                                                                                                                                                                                                                                                                       \label{inst:0129}\vfill
\and Konkoly Observatory, Research Centre for Astronomy and Earth Sciences, E\"{ o}tv\"{ o}s Lor{\'a}nd Research Network (ELKH), MTA Centre of Excellence, Konkoly Thege Mikl\'{o}s \'{u}t 15-17, 1121 Budapest, Hungary\relax                                                                                                                                     \label{inst:0136}\vfill
\and ELTE E\"{ o}tv\"{ o}s Lor\'{a}nd University, Institute of Physics, 1117, P\'{a}zm\'{a}ny P\'{e}ter s\'{e}t\'{a}ny 1A, Budapest, Hungary\relax                                                                                                                                                                                                                 \label{inst:0137}\vfill
\and Instituut voor Sterrenkunde, KU Leuven, Celestijnenlaan 200D, 3001 Leuven, Belgium\relax                                                                                                                                                                                                                                                                      \label{inst:0139}\vfill
\and Department of Astrophysics/IMAPP, Radboud University, P.O.Box 9010, 6500 GL Nijmegen, The Netherlands\relax                                                                                                                                                                                                                                                   \label{inst:0140}\vfill
\and University of Vienna, Department of Astrophysics, T\"{ u}rkenschanzstra{\ss}e 17, A1180 Vienna, Austria\relax                                                                                                                                                                                                                                                 \label{inst:0148}\vfill
\and Institute of Physics, Laboratory of Astrophysics, Ecole Polytechnique F\'ed\'erale de Lausanne (EPFL), Observatoire de Sauverny, 1290 Versoix, Switzerland\relax                                                                                                                                                                                              \label{inst:0150}\vfill
\and Kapteyn Astronomical Institute, University of Groningen, Landleven 12, 9747 AD Groningen, The Netherlands\relax                                                                                                                                                                                                                                               \label{inst:0156}\vfill
\and School of Physics and Astronomy / Space Park Leicester, University of Leicester, University Road, Leicester LE1 7RH, United Kingdom\relax                                                                                                                                                                                                                     \label{inst:0163}\vfill
\and Thales Services for CNES Centre Spatial de Toulouse, 18 avenue Edouard Belin, 31401 Toulouse Cedex 9, France\relax                                                                                                                                                                                                                                            \label{inst:0165}\vfill
\and Depto. Estad\'istica e Investigaci\'on Operativa. Universidad de C\'adiz, Avda. Rep\'ublica Saharaui s/n, 11510 Puerto Real, C\'adiz, Spain\relax                                                                                                                                                                                                             \label{inst:0169}\vfill
\and Center for Research and Exploration in Space Science and Technology, University of Maryland Baltimore County, 1000 Hilltop Circle, Baltimore MD, USA\relax                                                                                                                                                                                                    \label{inst:0171}\vfill
\and GSFC - Goddard Space Flight Center, Code 698, 8800 Greenbelt Rd, 20771 MD Greenbelt, United States\relax                                                                                                                                                                                                                                                      \label{inst:0172}\vfill
\and EURIX S.r.l., Corso Vittorio Emanuele II 61, 10128, Torino, Italy\relax                                                                                                                                                                                                                                                                                       \label{inst:0174}\vfill
\and Porter School of the Environment and Earth Sciences, Tel Aviv University, Tel Aviv 6997801, Israel\relax                                                                                                                                                                                                                                                      \label{inst:0175}\vfill
\and Harvard-Smithsonian Center for Astrophysics, 60 Garden St., MS 15, Cambridge, MA 02138, USA\relax                                                                                                                                                                                                                                                             \label{inst:0176}\vfill
\and ATOS for CNES Centre Spatial de Toulouse, 18 avenue Edouard Belin, 31401 Toulouse Cedex 9, France\relax                                                                                                                                                                                                                                                       \label{inst:0177}\vfill
\and HE Space Operations BV for European Space Agency (ESA), Camino bajo del Castillo, s/n, Urbanizacion Villafranca del Castillo, Villanueva de la Ca\~{n}ada, 28692 Madrid, Spain\relax                                                                                                                                                                          \label{inst:0179}\vfill
\and Instituto de Astrof\'{i}sica e Ci\^{e}ncias do Espa\c{c}o, Universidade do Porto, CAUP, Rua das Estrelas, PT4150-762 Porto, Portugal\relax                                                                                                                                                                                                                    \label{inst:0180}\vfill
\and LFCA/DAS,Universidad de Chile,CNRS,Casilla 36-D, Santiago, Chile\relax                                                                                                                                                                                                                                                                                        \label{inst:0182}\vfill
\and SISSA - Scuola Internazionale Superiore di Studi Avanzati, via Bonomea 265, 34136 Trieste, Italy\relax                                                                                                                                                                                                                                                        \label{inst:0186}\vfill
\and Telespazio for CNES Centre Spatial de Toulouse, 18 avenue Edouard Belin, 31401 Toulouse Cedex 9, France\relax                                                                                                                                                                                                                                                 \label{inst:0191}\vfill
\and University of Turin, Department of Computer Sciences, Corso Svizzera 185, 10149 Torino, Italy\relax                                                                                                                                                                                                                                                           \label{inst:0195}\vfill
\and Dpto. de Matem\'{a}tica Aplicada y Ciencias de la Computaci\'{o}n, Univ. de Cantabria, ETS Ingenieros de Caminos, Canales y Puertos, Avda. de los Castros s/n, 39005 Santander, Spain\relax                                                                                                                                                                   \label{inst:0198}\vfill
\and Centro de Astronom\'{i}a - CITEVA, Universidad de Antofagasta, Avenida Angamos 601, Antofagasta 1270300, Chile\relax                                                                                                                                                                                                                                          \label{inst:0208}\vfill
\and DLR Gesellschaft f\"{ u}r Raumfahrtanwendungen (GfR) mbH M\"{ u}nchener Stra{\ss}e 20 , 82234 We{\ss}ling\relax                                                                                                                                                                                                                                               \label{inst:0213}\vfill
\and Centre for Astrophysics Research, University of Hertfordshire, College Lane, AL10 9AB, Hatfield, United Kingdom\relax                                                                                                                                                                                                                                         \label{inst:0215}\vfill
\and University of Turin, Mathematical Department ""G.Peano"", Via Carlo Alberto 10, 10123 Torino, Italy\relax                                                                                                                                                                                                                                                     \label{inst:0222}\vfill
\and University of Antwerp, Onderzoeksgroep Toegepaste Wiskunde, Middelheimlaan 1, 2020 Antwerp, Belgium\relax                                                                                                                                                                                                                                                     \label{inst:0226}\vfill
\and INAF - Osservatorio Astronomico d'Abruzzo, Via Mentore Maggini, 64100 Teramo, Italy\relax                                                                                                                                                                                                                                                                     \label{inst:0229}\vfill
\and Instituto de Astronomia, Geof\`{i}sica e Ci\^{e}ncias Atmosf\'{e}ricas, Universidade de S\~{a}o Paulo, Rua do Mat\~{a}o, 1226, Cidade Universitaria, 05508-900 S\~{a}o Paulo, SP, Brazil\relax                                                                                                                                                                \label{inst:0232}\vfill
\and APAVE SUDEUROPE SAS for CNES Centre Spatial de Toulouse, 18 avenue Edouard Belin, 31401 Toulouse Cedex 9, France\relax                                                                                                                                                                                                                                        \label{inst:0239}\vfill
\and M\'{e}socentre de calcul de Franche-Comt\'{e}, Universit\'{e} de Franche-Comt\'{e}, 16 route de Gray, 25030 Besan\c{c}on Cedex, France\relax                                                                                                                                                                                                                  \label{inst:0242}\vfill
\and Theoretical Astrophysics, Division of Astronomy and Space Physics, Department of Physics and Astronomy, Uppsala University, Box 516, 751 20 Uppsala, Sweden\relax                                                                                                                                                                                             \label{inst:0246}\vfill
\and School of Physics and Astronomy, Tel Aviv University, Tel Aviv 6997801, Israel\relax                                                                                                                                                                                                                                                                          \label{inst:0251}\vfill
\and Astrophysics Research Centre, School of Mathematics and Physics, Queen's University Belfast, Belfast BT7 1NN, UK\relax                                                                                                                                                                                                                                        \label{inst:0253}\vfill
\and Centre de Donn\'{e}es Astronomique de Strasbourg, Strasbourg, France\relax                                                                                                                                                                                                                                                                                    \label{inst:0255}\vfill
\and Universit\'{e} C\^{o}te d'Azur, Observatoire de la C\^{o}te d'Azur, CNRS, Laboratoire G\'{e}oazur, Bd de l'Observatoire, CS 34229, 06304 Nice Cedex 4, France\relax                                                                                                                                                                                           \label{inst:0256}\vfill
\and Institute for Computational Cosmology, Department of Physics, Durham University, Durham DH1 3LE, UK\relax                                                                                                                                                                                                                                                     \label{inst:0261}\vfill
\and European Southern Observatory, Karl-Schwarzschild-Str. 2, 85748 Garching, Germany\relax                                                                                                                                                                                                                                                                       \label{inst:0262}\vfill
\and Max-Planck-Institut f\"{ u}r Astrophysik, Karl-Schwarzschild-Stra{\ss}e 1, 85748 Garching, Germany\relax                                                                                                                                                                                                                                                      \label{inst:0263}\vfill
\and Data Science and Big Data Lab, Pablo de Olavide University, 41013, Seville, Spain\relax                                                                                                                                                                                                                                                                       \label{inst:0267}\vfill
\and Barcelona Supercomputing Center (BSC), Pla\c{c}a Eusebi G\"{ u}ell 1-3, 08034-Barcelona, Spain\relax                                                                                                                                                                                                                                                          \label{inst:0275}\vfill
\and ETSE Telecomunicaci\'{o}n, Universidade de Vigo, Campus Lagoas-Marcosende, 36310 Vigo, Galicia, Spain\relax                                                                                                                                                                                                                                                   \label{inst:0280}\vfill
\and Asteroid Engineering Laboratory, Space Systems, Lule\aa{} University of Technology, Box 848, S-981 28 Kiruna, Sweden\relax                                                                                                                                                                                                                                    \label{inst:0284}\vfill
\and Vera C Rubin Observatory,  950 N. Cherry Avenue, Tucson, AZ 85719, USA\relax                                                                                                                                                                                                                                                                                  \label{inst:0289}\vfill
\and Department of Astrophysics, Astronomy and Mechanics, National and Kapodistrian University of Athens, Panepistimiopolis, Zografos, 15783 Athens, Greece\relax                                                                                                                                                                                                  \label{inst:0290}\vfill
\and TRUMPF Photonic Components GmbH, Lise-Meitner-Stra{\ss}e 13,  89081 Ulm, Germany\relax                                                                                                                                                                                                                                                                        \label{inst:0293}\vfill
\and IAC - Instituto de Astrofisica de Canarias, Via L\'{a}ctea s/n, 38200 La Laguna S.C., Tenerife, Spain\relax                                                                                                                                                                                                                                                   \label{inst:0298}\vfill
\and Department of Astrophysics, University of La Laguna, Via L\'{a}ctea s/n, 38200 La Laguna S.C., Tenerife, Spain\relax                                                                                                                                                                                                                                          \label{inst:0299}\vfill
\and Faculty of Aerospace Engineering, Delft University of Technology, Kluyverweg 1, 2629 HS Delft, The Netherlands\relax                                                                                                                                                                                                                                          \label{inst:0301}\vfill
\and Radagast Solutions\relax                                                                                                                                                                                                                                                                                                                                      \label{inst:0305}\vfill
\and Laboratoire Univers et Particules de Montpellier, CNRS Universit\'{e} Montpellier, Place Eug\`{e}ne Bataillon, CC72, 34095 Montpellier Cedex 05, France\relax                                                                                                                                                                                                 \label{inst:0306}\vfill
\and Universit\'{e} de Caen Normandie, C\^{o}te de Nacre Boulevard Mar\'{e}chal Juin, 14032 Caen, France\relax                                                                                                                                                                                                                                                     \label{inst:0313}\vfill
\and LESIA, Observatoire de Paris, Universit\'{e} PSL, CNRS, Sorbonne Universit\'{e}, Universit\'{e} de Paris, 5 Place Jules Janssen, 92190 Meudon, France\relax                                                                                                                                                                                                   \label{inst:0314}\vfill
\and SRON Netherlands Institute for Space Research, Niels Bohrweg 4, 2333 CA Leiden, The Netherlands\relax                                                                                                                                                                                                                                                         \label{inst:0324}\vfill
\and Astronomical Observatory, University of Warsaw,  Al. Ujazdowskie 4, 00-478 Warszawa, Poland\relax                                                                                                                                                                                                                                                             \label{inst:0325}\vfill
\and Scalian for CNES Centre Spatial de Toulouse, 18 avenue Edouard Belin, 31401 Toulouse Cedex 9, France\relax                                                                                                                                                                                                                                                    \label{inst:0327}\vfill
\and Universit\'{e} Rennes, CNRS, IPR (Institut de Physique de Rennes) - UMR 6251, 35000 Rennes, France\relax                                                                                                                                                                                                                                                      \label{inst:0333}\vfill
\and INAF - Osservatorio Astronomico di Capodimonte, Via Moiariello 16, 80131, Napoli, Italy\relax                                                                                                                                                                                                                                                                 \label{inst:0335}\vfill
\and Shanghai Astronomical Observatory, Chinese Academy of Sciences, 80 Nandan Road, Shanghai 200030, People's Republic of China\relax                                                                                                                                                                                                                             \label{inst:0338}\vfill
\and University of Chinese Academy of Sciences, No.19(A) Yuquan Road, Shijingshan District, Beijing 100049, People's Republic of China\relax                                                                                                                                                                                                                       \label{inst:0340}\vfill
\and Niels Bohr Institute, University of Copenhagen, Juliane Maries Vej 30, 2100 Copenhagen {\O}, Denmark\relax                                                                                                                                                                                                                                                    \label{inst:0343}\vfill
\and DXC Technology, Retortvej 8, 2500 Valby, Denmark\relax                                                                                                                                                                                                                                                                                                        \label{inst:0344}\vfill
\and Las Cumbres Observatory, 6740 Cortona Drive Suite 102, Goleta, CA 93117, USA\relax                                                                                                                                                                                                                                                                            \label{inst:0345}\vfill
\and CIGUS CITIC, Department of Nautical Sciences and Marine Engineering, University of A Coru\~{n}a, Paseo de Ronda 51, 15071, A Coru\~{n}a, Spain\relax                                                                                                                                                                                                          \label{inst:0354}\vfill
\and Astrophysics Research Institute, Liverpool John Moores University, 146 Brownlow Hill, Liverpool L3 5RF, United Kingdom\relax                                                                                                                                                                                                                                  \label{inst:0355}\vfill
\and IPAC, Mail Code 100-22, California Institute of Technology, 1200 E. California Blvd., Pasadena, CA 91125, USA\relax                                                                                                                                                                                                                                           \label{inst:0362}\vfill
\and IRAP, Universit\'{e} de Toulouse, CNRS, UPS, CNES, 9 Av. colonel Roche, BP 44346, 31028 Toulouse Cedex 4, France\relax                                                                                                                                                                                                                                        \label{inst:0363}\vfill
\and MTA CSFK Lend\"{ u}let Near-Field Cosmology Research Group, Konkoly Observatory, MTA Research Centre for Astronomy and Earth Sciences, Konkoly Thege Mikl\'{o}s \'{u}t 15-17, 1121 Budapest, Hungary\relax                                                                                                                                                    \label{inst:0380}\vfill
\and Departmento de F\'{i}sica de la Tierra y Astrof\'{i}sica, Universidad Complutense de Madrid, 28040 Madrid, Spain\relax                                                                                                                                                                                                                                        \label{inst:0396}\vfill
\and Ru{\dj}er Bo\v{s}kovi\'{c} Institute, Bijeni\v{c}ka cesta 54, 10000 Zagreb, Croatia\relax                                                                                                                                                                                                                                                                     \label{inst:0402}\vfill
\and Villanova University, Department of Astrophysics and Planetary Science, 800 E Lancaster Avenue, Villanova PA 19085, USA\relax                                                                                                                                                                                                                                 \label{inst:0418}\vfill
\and INAF - Osservatorio Astronomico di Brera, via E. Bianchi, 46, 23807 Merate (LC), Italy\relax                                                                                                                                                                                                                                                                  \label{inst:0424}\vfill
\and STFC, Rutherford Appleton Laboratory, Harwell, Didcot, OX11 0QX, United Kingdom\relax                                                                                                                                                                                                                                                                         \label{inst:0432}\vfill
\and Charles University, Faculty of Mathematics and Physics, Astronomical Institute of Charles University, V Holesovickach 2, 18000 Prague, Czech Republic\relax                                                                                                                                                                                                   \label{inst:0455}\vfill
\and Department of Particle Physics and Astrophysics, Weizmann Institute of Science, Rehovot 7610001, Israel\relax                                                                                                                                                                                                                                                 \label{inst:0466}\vfill
\and Department of Astrophysical Sciences, 4 Ivy Lane, Princeton University, Princeton NJ 08544, USA\relax                                                                                                                                                                                                                                                         \label{inst:0467}\vfill
\and Departamento de Astrof\'{i}sica, Centro de Astrobiolog\'{i}a (CSIC-INTA), ESA-ESAC. Camino Bajo del Castillo s/n. 28692 Villanueva de la Ca\~{n}ada, Madrid, Spain\relax                                                                                                                                                                                      \label{inst:0476}\vfill
\and naXys, University of Namur, Rempart de la Vierge, 5000 Namur, Belgium\relax                                                                                                                                                                                                                                                                                   \label{inst:0479}\vfill
\and CGI Deutschland B.V. \& Co. KG, Mornewegstr. 30, 64293 Darmstadt, Germany\relax                                                                                                                                                                                                                                                                               \label{inst:0487}\vfill
\and Institute of Global Health, University of Geneva\relax                                                                                                                                                                                                                                                                                                        \label{inst:0488}\vfill
\and Astronomical Observatory Institute, Faculty of Physics, Adam Mickiewicz University, Pozna\'{n}, Poland\relax                                                                                                                                                                                                                                                  \label{inst:0490}\vfill
\and H H Wills Physics Laboratory, University of Bristol, Tyndall Avenue, Bristol BS8 1TL, United Kingdom\relax                                                                                                                                                                                                                                                    \label{inst:0494}\vfill
\and Department of Physics and Astronomy G. Galilei, University of Padova, Vicolo dell'Osservatorio 3, 35122, Padova, Italy\relax                                                                                                                                                                                                                                  \label{inst:0501}\vfill
\and CERN, Geneva, Switzerland\relax                                                                                                                                                                                                                                                                                                                               \label{inst:0503}\vfill
\and Applied Physics Department, Universidade de Vigo, 36310 Vigo, Spain\relax                                                                                                                                                                                                                                                                                     \label{inst:0505}\vfill
\and Association of Universities for Research in Astronomy, 1331 Pennsylvania Ave. NW, Washington, DC 20004, USA\relax                                                                                                                                                                                                                                             \label{inst:0509}\vfill
\and European Southern Observatory, Alonso de C\'ordova 3107, Casilla 19, Santiago, Chile\relax                                                                                                                                                                                                                                                                    \label{inst:0517}\vfill
\and Sorbonne Universit\'{e}, CNRS, UMR7095, Institut d'Astrophysique de Paris, 98bis bd. Arago, 75014 Paris, France\relax                                                                                                                                                                                                                                         \label{inst:0522}\vfill
\and Faculty of Mathematics and Physics, University of Ljubljana, Jadranska ulica 19, 1000 Ljubljana, Slovenia\relax                                                                                                                                                                                                                                               \label{inst:0524}\vfill
}

\date{Received May 3, 2022 ; accepted June 1st, 2022 }

\abstract
{ We present the third data release of the European Space Agency's  \gaia\ mission, \gdrthree. This release includes a large variety of new data products, notably a much expanded radial velocity survey and a very extensive astrophysical characterisation of \gaia\ sources.}
{We  outline the content and the properties of \gdrthree, providing an overview of the main improvements in the data processing in comparison with previous data releases (where applicable) and a brief discussion  of the limitations of the data in this release.}
{The \gdrthree catalogue is the outcome of the processing of raw data collected with the \gaia\ instruments during the first 34 months of the mission by the Gaia Data Processing and Analysis Consortium.}
{The \gdrthree catalogue contains  the same source list, celestial positions,  proper motions, parallaxes, and  broad band photometry in the \gmag, \gbp,  and \grp pass-bands already present in the Early Third Data Release, \gedrthree. \gdrthree introduces an impressive wealth of new data products. More than 33 million objects in the ranges $\grvs < 14$ and  $3100 <\teff <14500 $, have new determinations of their mean radial velocities based on data collected by \gaia. We provide \grvs magnitudes for most sources with radial velocities, and  a line broadening parameter is listed for a subset of these. Mean \gaia\ spectra are made available to the community. The \gdrthree\ catalogue includes about 1 million mean spectra from the radial velocity spectrometer, and about 220 million low-resolution blue and red prism photometer \bprp mean spectra. The results of the analysis of epoch photometry are provided for some 10 million sources across 24 variability types. \gdrthree includes astrophysical parameters and source class probabilities for about 470 million and 1500 million sources, respectively, including stars, galaxies, and quasars. Orbital elements and trend parameters are provided for some $800\,000$ astrometric, spectroscopic and eclipsing binaries. More than $150\,000$  Solar System objects, including new discoveries, with preliminary orbital solutions and individual epoch observations are part of this release. Reflectance spectra derived from the epoch  \bprp spectral data are published for about 60\,000 asteroids. Finally, an additional data set is provided, namely the \gaia\ Andromeda Photometric Survey, consisting of the photometric time series for all sources located in a $5.5$ degree radius field centred on the Andromeda galaxy.}
{This data release represents a major advance with respect to \gdrtwo and \gedrthree because of the unprecedented quantity, quality, and variety of source astrophysical data. To date this is the largest collection of  all-sky spectrophotometry, radial velocities, variables, and astrophysical parameters derived from both low- and high-resolution spectra and includes a  spectrophotometric and dynamical survey of SSOs of the highest accuracy. The non-single star content surpasses  the existing data by orders of magnitude. The quasar host and galaxy light profile collection is the first such survey that is all sky and space based. The astrophysical information provided in \gdr{3} will unleash the full potential of \gaia's exquisite astrometric, photometric, and radial velocity surveys.}


\maketitle
\section{Introduction}
\label{sec:introduction}
The European Space Agency's (ESA) \gaia\ mission \citep{DR1-DPACP-18}, launched in 2013, is now at its third data release (\gdrthree). The aim of this paper is to  present the \gdrthree\ data products, providing a brief overview of the new features introduced in the processing and discussing the quality of the data. \gedrthree \citep{EDR3-DPACP-130} was the first instalment of the full \gdrthree\ and included astrometry and broad band photometry for a total of $1.8$ billion objects based on 34 months of satellite operations. Radial velocities for 7 million sources were copied over from the second data release, \gdrtwo, where a small number of spurious radial velocities were removed \citep{EDR3-DPACP-121}.

\gdrthree complements \gedrthree, introducing a vast array of new data products based on the same source catalogue and raw observations at the basis of \gedrthree. Indeed, the astrometry and broad band photometry in \gmag, \bpmag, and \rpmag from \gedrthree are repeated in the new catalogue (but see section~\ref{ssec:gcorr} for \gmag  correction). New data products in \gdrthree include mean low-resolution blue and red photometer (\bprp) spectra and high-resolution Radial Velocity Spectrometer (RVS) spectra; new estimates of mean radial velocities to a fainter limiting magnitude; line broadening and chemical composition information derived from RVS spectra; variable-star classification and characterisation; and photometric time series for over 20 classes of variables. Photometric time series for all sources, variable and non-variable, are available for a field centred on the Andromeda galaxy. A large sample of Solar System objects (SSOs) with orbital solutions and epoch observations is part of the data release, together with  reflectance spectra for a subset of those. \gdrthree also includes results for non-single stars (NSS), quasars, and galaxies. For a large fraction of the objects, the catalogue lists astrophysical parameters (APs) determined from parallaxes, broad band photometry, and the mean RVS or mean \bprp\ spectra. 

To enhance the scientific exploitation of the data, the \gaia\ archive includes pre-computed cross-matches with selected  external optical and near-infrared photometric and spectroscopic surveys. In addition to the catalogues described in \citet{EDR3-DPACP-130}, \gdrthree includes the sixth data release from the Radial Velocity Experiment (RAVE) survey \citep{2020AJ....160...83S}.
The \gaia\ Universe Model Snapshot \citep[GUMS, version 20,][]{2012A&A...543A.100R} and the corresponding simulated Gaia catalogue (GOG) were also already included in \gedrthree. The details on these supplementary data can be found in the online documentation\footnote{\gdr{3} online documentation: \url{https://www.cosmos.esa.int/web/gaia-users/archive/gdr3-documentation}}. 

This paper is organised as follows. Section~\ref{sec:dataprocessing} summarises the \gaia\ instruments and their wavelength ranges; section~\ref{ssec:edr3} summarises the properties of the astrometric and photometric data already included in \gedrthree; section~\ref{ssec:photspectra} presents the \bprp\ spectra; section~\ref{ssec:vrad} outlines the RVS data processing, listing the improvements implemented in \gdrthree, and presents the new RVS data products, the line broadening velocity (\vbroad) and the \grvs magnitude. Section~\ref{ssec:vari} deals with the variable source content of the catalogue including the \gaia Andromeda Photometric Survey (GAPS); section~\ref{ssec:binary} discusses the NSS content of the release; section~\ref{ssec:solar} comments on the SSOs; section~\ref{ssec:classif} provides information about the APs; section~\ref{ssec:extended} presents the extended object (EO) data processing and results. Section~\ref{ssec:QSOs} describes the extragalactic content of \gdrthree, that is, information on galaxies and  quasi-stellar objects (QSOs) or `quasars' derived by several data processing modules. Section~\ref{ssec:science} comments on the quality of the release and section~\ref{ssec:tools} provides references to the software tools that are offered to the users to deal with \gdrthree data. Finally, section~\ref{ssec:conclusions} presents some concluding remarks.

Each data product section briefly summarises the main limitations of the data, and makes reference to the relevant \gaia\ Collaboration and \gaia\ Data Processing and Analysis Consortium (DPAC) papers where more details can be found. For a number of technical details, we refer the reader to the online documentation. 

All the papers accompanying \gdrthree are published in the Astronomy \& Astrophysics special issue on \gdrthree. Finally, we recall that all the {\gaia} data releases are made  available through the archive hosted by ESA\footnote{https://archives.esac.esa.int/gaia}, as described in \citet{EDR3-DPACP-130} and references therein. Partner and affiliated data centres in Europe, the United States, Japan, Australia, and South Africa  provide access to the data through their own facilities. 

\section{Data processing}
\label{sec:dataprocessing}
The \gaia\ satellite has three main instruments on board: the astrometric instrument collecting images in \gaia’s white-light G-band (330–1050 nm), the blue BP and red RP prism photometers for low-resolution spectra, and, finally, the RVS. \bprp spectral data cover the wavelength ranges 330--680~nm and 640--1050~nm, respectively. The resolution is variable and ranges from 30 to 100 for BP and 70 to 100 for RP in $\lambda/{\Delta\lambda}$, depending on the position in the spectrum and on the CCD \citep[see][]{EDR3-DPACP-119}.   
In the RVS spectra, the starlight is dispersed over about 1100
pixels in the \gaia\ telescopes scanning direction (along-scan, AL), sampling the wavelength
range from 845 to 872 nm. The resolving power is R $\sim$ 11\,500
(with a resolution element of about 3 pixels). As the wings of the spectra are excluded from the processing, the effective wavelength range of
the processed spectra is reduced to 846--870~nm \citep{CU6-DR3-documentation}. 
Similarly to \gedrthree, \gdr{3} is based on data collected over a 34-month time interval \citep[for details see][]{EDR3-DPACP-130}.
The \gaia\ data processing is the responsibility  of DPAC and is presented in \citet{DR1-DPACP-18}.

\begin{figure*}
    \sidecaption
    \includegraphics[width=12cm]{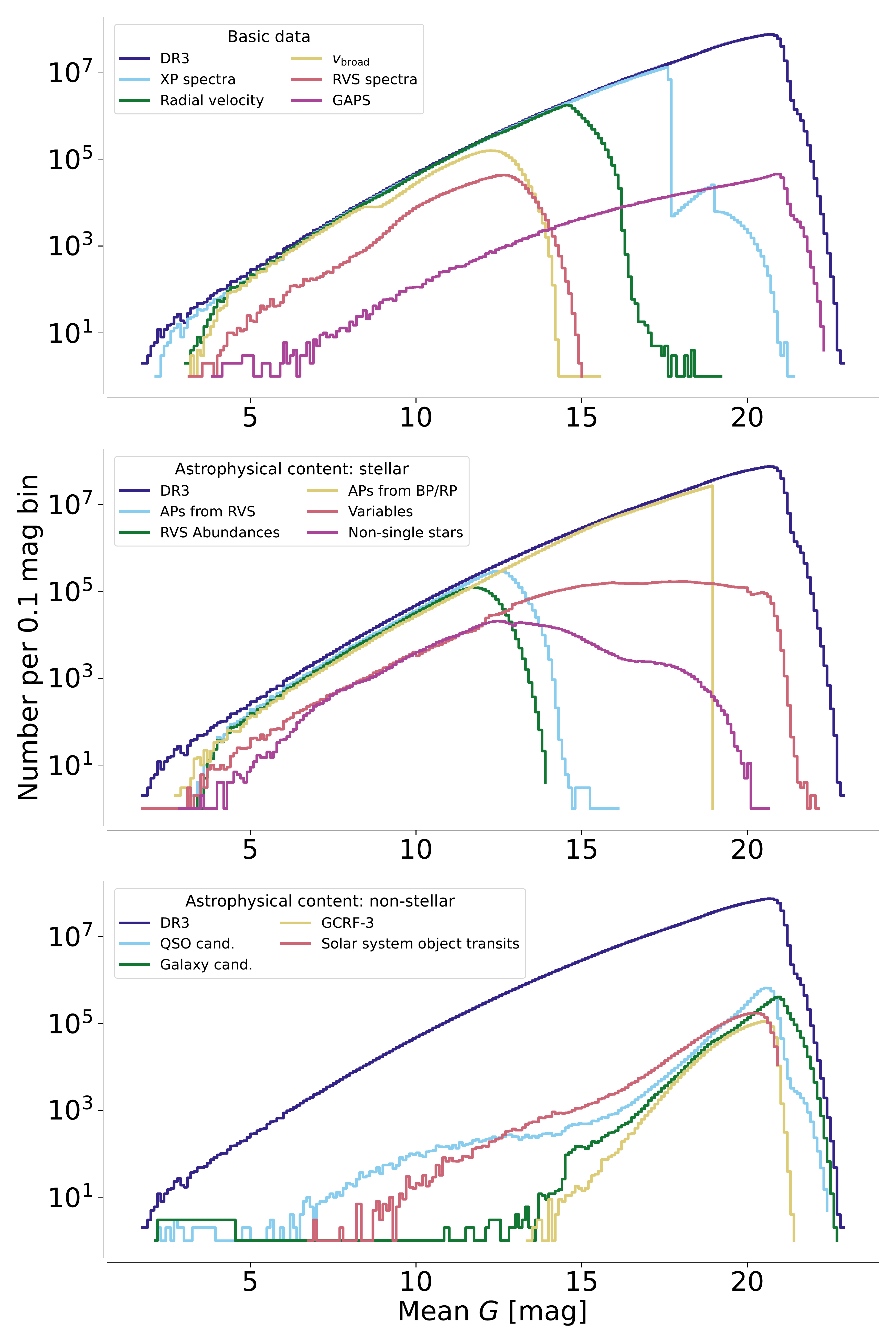}
    \caption{Distribution of the mean values of $G$ for the main \gdr{3} components shown as histograms with bins of $0.1$ mag in width. The top panel shows the histograms for the basic observational data in \gdr{3} (spectra, radial velocities, \vbroad, photometric time series). The middle panel shows histograms for the stellar astrophysical contents, and the bottom panel shows the non-stellar astrophysical contents. The sharp transitions in the top and middle panels at $G=17.65$ and $G=19$ are caused by the limit on the brightness of sources for which \bprp spectra are published and the limit up to which astrophysical parameters were estimated. The SSO histogram shows the distribution of the transit-level $G$-band magnitudes \citep[see][for a distribution of solar system objects in absolute magnitude $H$]{DR3-DPACP-150}. The QSO and galaxy candidate histograms extend to very bright magnitudes which is a consequence of favouring completeness over purity in these samples, and not applying any filtering to remove them \citep[see][]{DR3-DPACP-101}.}
    \label{fig:gmaghistos}
\end{figure*}

The basic statistics on the source numbers for each of the data products in \gdr{3} can be found in \tabref{tab:dr3stats}, with further details in \tabref{tab:dr3statsdetail}.  The categories listed in the tables are
described in the text below. A visual impression of the release contents is given in \figref{fig:gmaghistos} which shows histograms of the distribution in \gmag of the main categories of data products in \gdr{3}.

\begin{table*}[t!]
  \caption{Number of sources of a certain type, or the number of sources for which a given data product is available in \gdr{3}.}
  \label{tab:dr3stats}
  \centering
  \begin{tabular}{>{\raggedright}p{0.35\linewidth}r>{\raggedright\arraybackslash}p{0.3\linewidth}}
    \hline\hline
    Data product or source type & Number of sources & Comments\\
    \hline
    Total & \gdrthreetotal \\
    5-parameter astrometry & \gdrthreefiveptot \\
    6-parameter astrometry & \gdrthreesixptot \\
    2-parameter astrometry & \gdrthreetwoptot \\
    \gaia-CRF3 sources & \gdrthreegcrftot & \\
    ICRF3 sources used for frame orientation &  \gdrthreeicrfused & \\
    \gaia-CRF3 sources used for frame spin & \gdrthreespinused & \\
    $G$-band & \gdrthreewithgtot \\
    \gbp-band & \gdrthreewithbptot \\
    \grp-band & \gdrthreewithrptot \\
    Photometric time series & \gdrthreephottimeseries \\
    Gaia Andromeda Photometric Survey & \gdrthreegaps & Photometric time series for all sources in $5.5^\circ$ radius field around M31\\
    Radial velocity & \gdrthreevradtot & $\grvs<14$, $3100<\teff<14\,500$~K\\
    \grvs-band & \gdrthreegrvstot \\
    $v_\mathrm{broad}$ & \gdrthreevbroadtot & spectral line broadening parameter \\
    Radial velocity time series & \gdrthreeepochrvtot & Sample of Cepheids and RR~Lyrae \\
    BP/RP mean spectra & \gdrthreexpspectratot & $\gmag<17.65$ with small number of exceptions\\
    RVS mean spectra & \gdrthreervsspectratot & AFGK spectral types with $\mathrm{SNR}>20$, and sample of lower SNR spectra \\
    Variable sources & \gdrthreevaritot & See \tabref{tab:dr3statsdetail} \\
    Object  DSC classification  & \gdrthreedsc & \\
     Self-organised map of poorly classified sources & 1 & $30\times30$ map and prototype spectra\\
    APs from mean BP/RP spectra & \gdrthreeapsfromxp & $\gmag<19$, see \tabref{tab:dr3statsdetail} \\
    APs from mean RVS spectra & \gdrthreeapsfromrvs & \\
    Chemical abundances from mean RVS spectra &\gdrthreeabunfromrvs  & Up to 12 elements \\
    DIBs from mean RVS spectra & \gdrthreedibs & \\
    Non-single stars & \gdrthreenss & astrometric, spectroscopic, eclipsing, orbits, trends, see
    \tabref{tab:dr3statsdetail} \\
    QSO candidates & \gdrthreeqsocandidates & High completeness, low purity \\
    QSO redshifts & \gdrthreeqsoredshifts & \\
    QSO host galaxy detected & \gdrthreeqsohosts & \\
    QSO host galaxy profile & \gdrthreeqsohostprofile & \\
    Galaxy candidates & \gdrthreegalaxycandidates & High completeness, low purity \\
    Galaxy redshifts & \gdrthreegalaxyredshifts & \\
    Galaxy profiles & \gdrthreegalaxyprofile & \\
    Solar System objects & \gdrthreesso & Epoch astrometry and photometry \\
    SSO reflectance spectra & \gdrthreessoreflectance & \\
    Total galactic extinction maps & 5 & HEALPix levels 6--9, and optimum HEALPix level \\
    Science alerts & \gdrthreealerts & Triggered in the period 25-07-2014 to 28-05-2017 \\
    \hline
  \end{tabular}
\end{table*}

\begin{table*}[t!]
  \caption{Further details on the number of sources of a certain type, or the number of sources for which a given data product is available in \gdr{3}.}
  \label{tab:dr3statsdetail}
  \centering
  \begin{tabular}{>{\raggedright}p{0.4\linewidth}r>{\raggedright\arraybackslash}p{0.3\linewidth}}
    \hline\hline
    Data product or source type & Number of sources & Comments\\
    \hline
    \multicolumn{3}{c}{Variable sources} \\
    \hline
    Total & \gdrthreevaritot & \\ 
    Classified with supervised machine learning & \gdrthreevarimlclassified & 24 variability types or type groups \\
    Active galactic nuclei & \gdrthreevariagn & \\
    Cepheids & \gdrthreevaricepheids & \\
    Compact companions & \gdrthreevaricompact & \\
    Eclipsing binaries & \gdrthreevarieclipsing & \\
    Long-period variables & \gdrthreevarilpv & \\
    Microlensing events & \gdrthreevarimulens & \\
    Planetary transits & \gdrthreevariplanets & \\
    RR~Lyrae stars & \gdrthreevarirrl & \\
    Short-timescale variables & \gdrthreevarishortts & \\
    Solar-like rotational modulation variables & \gdrthreevarisolarlike & \\
    Upper-main-sequence oscillators & \gdrthreevariums & \\
    \hline
    \multicolumn{3}{c}{Astrophysical parameters from mean BP/RP spectra} \\
    \hline
    Total & \gdrthreeapsfromxp & $\gmag<19$ \\
    Spectroscopic parameters & \gdrthreeapsfromxp \\
    Interstellar extinction and distances & \gdrthreeapsfromxp \\
    MCMC samples from the BP/RP AP estimation & \gdrthreemcmcgspphot \\
    APs assuming an unresolved binary & \gdrthreeapsbinary & \\
    MCMC samples from BP/RP unresolved binary AP estimation & \gdrthreemcmcmsc \\
    Evolutionary parameters & \gdrthreeevolpars & mass, age, evolutionary stage \\
    Stars with emission-line classifications & \gdrthreexpels & \\
    Sources with spectral types & \gdrthreexpspt & \\
    Hot stars with spectroscopic parameters & \gdrthreexphotstars & \\
    Ultra-cool stars & \gdrthreexpucd & \\
    Cool stars with activity index &\gdrthreexpcoolactive  & \\
    Sources with H$\alpha$ emission measurements & \gdrthreexphalpha & \\
    \hline
    \multicolumn{3}{c}{Non-single stars} \\
    \hline
    Total & \gdrthreenss & \\
    Acceleration solutions & \gdrthreenssacceleration & \\
    Orbital astrometric solutions & \gdrthreenssorbitalastromcomb & including astroSpectroSB1 combined solutions \\
    Orbital spectroscopic solutions (SB1/SB2) & \gdrthreenssorbitalspectrocomb & including astroSpectroSB1 combined solutions\\
    Trend spectroscopic solutions &\gdrthreensstrendspectro  \\
    Eclipsing binaries & \gdrthreensseclipsing& including eclipsingSpectro combined solutions \\
    \hline
  \end{tabular}
\end{table*}

\section{Astrometry and broad band photometry}\label{ssec:edr3}

The astrometric and broad band photometry content of \gdrthree is  the same as that for \gedrthree, except for the addition of the \grvs photometry, but for convenience we summarise here the properties of these data products. \gedr{3} provided celestial positions and the apparent brightness in $G$ for $1.8$ billion sources. For $1.5$ billion of those sources, parallaxes, proper motions, and the \bpminrp colour were also published. \gdr{3} therefore contains some 585 million sources with five-parameter astrometry (two positions, the parallax, and two proper motion components), and about 882 million sources with six-parameter (6-p) astrometry, including an additional pseudo-colour parameter. We refer to \cite{EDR3-DPACP-130} for details on \gedrthree, a summary of the astrometric and photometric data processing, and a  summary of the limitations of these data and guidance on their use. Detailed descriptions of the astrometry in \gedrthree (and thus also \gdrthree) are provided in \cite{EDR3-DPACP-128}, while the broad band photometry is described in detail in \cite{EDR3-DPACP-117}. Details on the construction of the \gdr{3} source list can be found in \cite{EDR3-DPACP-124}, while the basic inputs to the astrometric and $G$-band photometric processing (the $G$-band source image positions and fluxes) are described in \cite{EDR3-DPACP-73}. The validation of the astrometry and broad band photometry is summarised in \cite{EDR3-DPACP-126}. The photometric pass-bands for \gmag, \gbp, and \grp are provided in \cite{EDR3-DPACP-117}, and the one for \grvs in \cite{DR3-DPACP-155}.

Finally, the \gaia Science Alerts project and content are described in \citet{Hodgkin2021}.

\subsection{$G$-band photometry for 6-p and 2-p sources}\label{ssec:gcorr}

Section 7.2 of \cite{EDR3-DPACP-130} and Section 8.3 of \cite{EDR3-DPACP-117} describe a correction to be applied to the $G$-band photometry for sources with 6-p and 2-p astrometric solutions. This correction was provided in the form of Python code and Astronomical Data Query Language (ADQL) recipes. \textit{We note here that these corrections are included in \gdr{3} and therefore should  not be applied when working with photometry extracted from the \gdr{3} data tables in the \gaia archive.}

As noted in \cite{EDR3-DPACP-130} the \gmag-band photometry for a small number of sources is not listed in \gedr{3}. This issue has not been fixed for \gdr{3}. The magnitudes can be found in a separate table to be provided through the \gdr{3} `known issues' web pages \citep[for details see section 8.2 in][]{EDR3-DPACP-117}.

\subsection{Celestial reference frame}

Given that the astrometry in \gdr{3} is unchanged from
\gedr{3}, it follows that the source positions and proper motions
are provided with respect to the \gaia-CRF3, the third
realisation of the Gaia celestial reference frame. \gaia-CRF3
is aligned with the third realisation of the International Celestial
Reference Frame in the radio (ICRF3) to about $0.01$ mas root-mean-square
(RMS) at epoch J2016.0 (barycentric coordinate time, TCB), and globally
non-rotating with respect to quasars to within $0.005$~\maspyr\  RMS. The
\gaia-CRF3 is defined by the positions and proper motions of
\gdrthreegcrftot QSO-like sources that were selected using \gedr{3}
astrometry. For the alignment and spin of the \gaia-CRF3, the special
sets of 2007 and 428\,034 sources, respectively, were used based on the
preliminary astrometric solutions known as AGIS3.1
\citep{EDR3-DPACP-128}. The construction and properties of the
\gaia-CRF3, the
comparison to the ICRF3, and the procedure to fix the alignment
and spin of the astrometric solution are described in \citet{EDR3-DPACP-133}.

\subsection{Systematic errors}

The systematic errors present in the astrometry and broad band photometry published in \gedr{3} carry over to \gdr{3}. The conclusions from investigations during the data processing for \gedr{3} astrometry were that the global parallax bias for \gdr{3}, as measured from quasars, is $-17$~\muas. The RMS angular (i.e.\ source to source) covariances of the parallaxes and proper motions on small scales are $\sim26$~\muas and $\sim33$~\muasyr, respectively \citep[see][for details]{EDR3-DPACP-128}. The parallax bias (and the proper motion systematic errors) varies as a function of magnitude, colour, and celestial position. This is extensively investigated in \cite{EDR3-DPACP-132} and a recipe for correcting the parallaxes is given. The systematic errors in the broad band photometry are described in \cite{EDR3-DPACP-117}.

Since the release of \gedr{3} several investigations of the systematic errors have been published.  The parallax bias was investigated for specific sets of sources by \cite{2021ApJ...907L..33S}, and \cite{2021ApJ...911L..20R} (eclipsing binaries); \cite{2021ApJ...910L...5H} (Red Clump stars); \cite{2021AJ....161..214Z} (red giant branch stars with asteroseismic parallaxes); \cite{2022AJ....163..149W} (red giant stars), \cite{2022MNRAS.509.4276F} (stellar clusters); \cite{2021A&A...654A..20G}, \cite{2021A&A...653A..61K}, \cite{2021ApJ...908L...6R}, and \cite{2022MNRAS.509.2566M} (Cepheids and RR~Lyrae stars); \cite{2022A&A...657A.130M}, and \cite{2021A&A...649A..13M} (Magellanic Clouds and globular clusters). The small-scale covariances are investigated in \citet{2021AJ....161..214Z}, \citet{2021MNRAS.505.5978V}, and  \citet{2021A&A...649A..13M}, while assessments of the parallax uncertainties can be found in \cite{2021MNRAS.506.2269E},  \cite{2021MNRAS.505.5978V}, and \cite{2021A&A...649A..13M}.

The systematic errors in the proper motions and positions of stars in \gedr{3} were investigated by \cite{2021A&A...649A.124C} and \cite{2021EGUGA..23.8604L}. \cite{2021A&A...649A.124C} demonstrated that the proper motions of bright ($G\lesssim13$) stars show  a residual spin with respect to the fainter stars by up to $80$~\muasyr, and they provide a recipe to correct for this effect. \cite{2021EGUGA..23.8604L} found differences in the alignment to the ICRF3 between bright and faint sources of about $0.5$~mas.  Systematic errors in the broad band photometry were investigated by \cite{2021ApJ...908L..14N}, \cite{2021ApJ...908L..24Y}, and \cite{2021MNRAS.505.5941T}.

We stress here that including the above references does not imply an endorsement by the \gaia Collaboration of the results or any systematic error correction recipes provided in those papers. Nevertheless, community efforts to investigate the quality of the \gaia data are highly appreciated and have in several cases led to an improvement in the validation procedures used during the \gaia data processing.

\section{\bprp spectra}\label{ssec:photspectra}
 
\bprp  spectral observations are  transmitted to the ground from the satellite in small windows  surrounding the position of the source. The size of the \bprp windows is 60 pixels AL by 12 pixels in the across scan (AC) direction, corresponding to an area in the sky of approximately 3.5\arcsec\ by 2.1\arcsec. The size of the window affects the detection of sources in crowded regions, resulting in  partially overlapping windows. These windows are excluded from the \gdrthree data processing. A special treatment will be implemented in future data releases.

\subsection{Data processing}\label{ssec:bprpproc} 
The \bprp processing is described in detail in \citet{EDR3-DPACP-118} and \citet{EDR3-DPACP-120}.
The spectra are first calibrated to an internal reference instrument which is homogeneous across all devices, observing configurations, and time. This is achieved purely from \bprp data for a sufficiently large subset of sources selected to cover all calibration units (windowing strategies, gates, magnitude ranges, time).  The internal calibration removes a number of  effects such as bias, background, geometry, differential dispersion, and variations in response and in the line spread function (LSF)  across the focal plane \citep{EDR3-DPACP-119, EDR3-DPACP-118}.
The internal reference system is linked to the absolute system (both in terms of flux and wavelength) via the external calibration, which is based on a dedicated catalogue of spectro-photometric calibrators \citep{2021MNRAS.503.3660P}.  More details concerning  the calibration of the \bprp spectral data to the absolute reference system are presented  in \citet{EDR3-DPACP-120}.  

We release mean low-resolution BP/RP spectra for about 220 million sources. They are selected to have a reasonable number of observations (more than 15 CCD transits) and to be
sufficiently bright to ensure good signal-to-noise ratio (S/N), that is, \gmag$<17.65$. To this list, we add a few samples of specific objects that could be as faint as \gmag$\sim 21.43$: about 500
sources used for the calibration of the BP/RP data, a catalogue of
about 100\,000  white dwarf candidates, 17\,000 galaxies, about 100\,000 QSOs,
about 19\,000 ultra-cool dwarfs, 900 objects that were considered
to be  representative for each of
the 900 neurons of the self-organising maps (SOMs) used by the outlier
analysis (OA) module (see section \ref{ssec:classif})  and  19 solar
analogues \citep{EDR3-DPACP-118}. 

The S/N of the \bprp spectra varies depending on the magnitude and colour of the source. In the range $9<$\gmag$<12,$ it can reach 1000 in the central part of the RP spectral range and it is of the order of 100 at \gmag$\sim$~15.
These spectra have been extensively used as input for further data processing inside DPAC (see e.g. sections~\ref{ssec:solar},  \ref{ssec:classif}, \ref{ssec:science}), which provides a strong validation of their exceptional quality.

\subsection{Data representation}
The source mean spectra are provided in a continuous representation: they are described by an array of coefficients to be applied to a set of basis functions. We use a set of 55 basis functions for BP and  55  basis functions for RP (referred to as `bases') defined as a linear combination of Hermite functions \citep{EDR3-DPACP-119, EDR3-DPACP-118}. 
In low-S/N spectra, as for instance  at faint magnitudes, it is possible that higher order bases  are over-fitting the noise in the observed data. In particular, low-S/N spectra when sampled in pseudo-wavelength can exhibit unrealistic features (wiggles).
To mitigate this problem,
\citet{EDR3-DPACP-119} suggest a statistical criterion to select the coefficients that can be dropped without losing information (truncation). Non-truncated and truncated spectra are in agreement within the noise. 
It should be noted that sharp features in the spectra, such as emission lines, can only be reproduced using higher order bases and therefore imply a larger number of significant coefficients. As the rejection criterion is statistical, it might happen that too few or too many coefficients are removed. This might affect faint objects with sharp spectral features, such as QSOs, or emission line stars.  The effects of the truncation for various object classes are described in Sect. 3.4.3 of \citet{EDR3-DPACP-118}.

To allow the users to decide how many coefficients are relevant for their scientific case,  all the 55 coefficients of the basis functions are released. The number of coefficients returned by the truncation criterion  is given in the parameters \textit{bp\_n\_relevant\_bases}
and \textit{rp\_np\_relevant\_bases}
available in the \textit{xp\_summary}\footnote{\phottable{xp_summary}} 
table and
in the mean continuous spectra available via Datalink\footnote{Datalink is an IVOA protocol that specifies a service container which can represent or accommodate a variety of services. Tutorials on the access to \gaia\ Datalink products can be found at {\url{https://www.cosmos.esa.int/web/gaia-users/archive/ancillary-data}}}.

In addition to the continuous representation,  sampled  spectra representation (i.e.\ in the form of integrated flux vs.\ pixel)  in both internal and absolute flux can be calculated  using the Python package GaiaXPy  made available with \gdrthree (see section~\ref{ssec:tools}). Sampling the spectra on a discrete grid in pseudo-wavelengths or absolute wavelengths results in a loss of information. In particular, the full covariance matrix is provided in the continuous representation, whereas it cannot be calculated in a spectrum sampled  on a grid with more points than the number of basis function coefficients. Although sampled spectra are made available for a subset of sources in \gdr{3}, we strongly encourage users of the \bprp\ spectra to make use of the continuous representation to  maximise the scientific use of these data.

\section{RVS data products}\label{ssec:vrad}

\gdrtwo was the first release to include RVS radial velocities based on 22 months of data for stars at $\grvs \le 12$ and with effective temperatures $3500 < \teff < 6900 $ K.  

\gdrthree contains newly determined radial velocities for about $33.8$ million stars with $\grvs \le 14$ and with  $3100 \le \teff \le 14500$~K. Additional data products  are published for the first time: \vbroad for about $3.5$ million stars, \grvs magnitudes for more than 32 million stars, mean spectra for slightly less than 1 million stars, and epoch radial velocities  for about 1000 RR~Lyrae and roughly 800 Cepheids of different types \citep{DR3-DPACP-169}.
All these data come with quality parameters.
Users are advised to  treat the \gdrthree radial velocity catalogue as completely independent of \gdrtwo.

The spectroscopic pipeline and the improvements since \gdrtwo are described in Chapter 6 of the online documentation  \citep{CU6-DR3-documentation}. All the products of the spectroscopic pipeline are available in the table  \textit{gaia\_source}\footnote{\gstable{gaia_source}}, 
except  for the mean spectra (see  section \ref{ssec:rvsSpectra}).
The transit radial velocities for the Cepheids and RR~Lyrae stars are published in the table \textit{vari\_epoch\_radial\_velocity}\footnote{\varitable{vari_epoch_radial_velocity}}.

 The products of the RVS pipeline, their properties, and their validation are described in more detail in dedicated papers: the radial velocity determination is described in \citet{DR3-DPACP-159}, the specific treatments to measure the hot star radial velocities in \citet{DR3-DPACP-151}, and the radial velocity processing of the double-lined spectra is presented in  \citet{DR3-DPACP-161}.  The \vbroad  determination is discussed in \citet{DR3-DPACP-149} and the \grvs magnitudes and the RVS pass-band in \citet{DR3-DPACP-155}.
 
For each valid RVS spectrum  entering the pipeline, the transit radial velocity is computed through a fit of the RVS spectrum relative to an appropriate synthetic template spectrum. In \gdrthree, for stars  cooler than 7000~K, the template input parameters are mostly taken from intermediate results of Apsis \citep{DR3-DPACP-157}  based on \gdrtwo \bprp spectra (see section~\ref{ssec:classif}). For  hotter stars, they were derived as explained in \citet{DR3-DPACP-151}.

\subsection{ Data processing improvements}

A number of  improvements were implemented in the \gdrthree RVS pipeline. Here we list the most significant.  Bright stars are processed using the method already implemented  in \gdrtwo, that is, the combined radial velocity for stars at \grvs $< 12$ is the median of the single-transit radial velocities.
However, this method is not very efficient when the S/N is low. The radial velocities of  faint stars in the range  $12< \grvs<14 $ mag are obtained from the averaged  single-transit cross-correlation functions referring to the Solar System barycentre.

To improve performance at the faint end, we implemented an improved stray-light-correction procedure. The correction map is  estimated every 30 hours from the faint-star spectra edges and from the virtual objects\footnote{Virtual objects are empty windows acquired on a predefined pattern for calibration purposes}, while  in \gdrtwo a single stray-light map was computed from  the Ecliptic Pole scanning law\footnote{During the early weeks of the mission, the \gaia\  spin axis followed the Sun on the ecliptic, scanning the North and South Ecliptic Poles every six hours  \citep[see][Section 5.2]{DR1-DPACP-18}} data.

We introduce a deblending procedure when the transit spectra are contaminated by nearby sources falling inside the  RVS window. These deblended spectra are then used to obtain radial velocities and mean spectra,  while in \gdrtwo they were simply removed from the pipeline. This allows us to process a  larger number of epoch spectra per source, increasing the S/N.   However, we point out  that only the clean non-blended transits are used  to derive \grvs, and \vbroad. 

The LSF was calibrated in both the AL and AC  directions. The LSF-AL calibration has reduced the systematic shifts between the two fields of view affecting the wavelength calibration zero point and the epoch radial velocities. The LSF-AC calibration  was also used in the deblending procedure and in the estimation of the flux lost outside the window for estimation of \grvs. 

\vbroad is computed for each transit, excluding deblended spectra, and then averaged. In addition to the projected rotational velocity, $v\sin i$, \vbroad can include other physical effects like macro-turbulence,  residual instrumental effects (LSF model uncertainty), and template mismatches.

We consider that a spectrum can be  contaminated by nearby sources  with $\grvs<15$ even if they are not located inside the RVS window. In that case, the target spectrum is removed when the contaminant differs from the target source by less than 3 mag. This effect was neglected in \gdrtwo.

Finally, \grvs (\textit{grvs\_mag}  in \textit{gaia\_source})
is calculated as the median of the single-transit \grvs measurements. Values of \grvs fainter than $14.1$ were regarded as  spurious and removed because they could have been caused by an inaccurate background estimate.

\subsection{RVS spectra} \label{ssec:rvsSpectra}


The published mean RVS spectra  are identified using the column \textit{has\_rvs}  in the \textit{gaia\_source}
table. Their spectra are available through the Datalink interface in the table  \textit{rvs\_mean\_spectrum}\footnote{\phottable{rvs_mean_spectrum}}. The mean spectra and their processing are described in detail in \citet{DR3-DPACP-154}. In summary,  the transit RVS spectra are extracted, cleaned, deblended (if needed), wavelength calibrated and normalised either to their pseudo-continuum or by scaling with a constant (the latter for cool stars or noisy spectra). The spectra are then shifted to the rest frame using the epoch radial velocities for the bright stars for \grvs $\le 12$ mag, or using the combined radial velocity for the faint stars; they are then interpolated into a common wavelength array spanning 846--870~nm with a step of 0.01 nm and averaged. S/N information is also provided in the table \textit{gaia\_source}.
  
The released RVS spectra are selected from among stars of spectral type AFGK with S/N $>20$. In addition, a sample of low-S/N spectra spanning all spectral types is added \citep[see][]{DR3-DPACP-154}. A known issue is that the published spectra are not uniformly distributed over the sky.

\subsection{Caveats}

The contamination of the flux of an RVS target source by the flux of a nearby bright source can produce a spurious population of stars with high radial velocities. This issue was known from \gdrtwo \citep{EDR3-DPACP-121}. In addition bright-star contamination produces  biased radial velocities which do not necessarily stand out from the overall radial velocity distribution when two sources  are separated by $1.8\arcsec$ in the direction perpendicular to the scan. These stars were removed from the catalogue \citep{DR3-DPACP-159}.

Radial velocity uncertainties are generally very small, of the order of a few \kms\ or even less than 1~\kms\ at the bright end. However, they are   slightly  underestimated for bright stars. A correction is proposed in \citet{DR3-DPACP-127}.

\section{Variables}\label{ssec:vari}

The approach to the variability analysis in  successive data releases is iterative, including at each data release  more variability types with lower signal to noise. In \gdrthree we publish a total of  more than 10 million variable sources  in about 24 variability types (and their time series), in addition to  approximately 2.5~million galaxies. 
The two-dimensional structure of galaxies is observed  by \gaia  over a range of position angles. This can induce  spurious (non-intrinsic) photometric variability. This effect is used to identify galaxies, but their time series are not released.

The \gdrthree variability content is a great leap in comparison with \gdrtwo where we reached more than 550\,000 stars, with six variability types.

In the variability pipeline (VARI) processing, the variability was first tested in the time domain, and was then  characterised in the Fourier domain and classified by multiple classifiers. We refer to the documentation and to \citet[][and references therein]{DR3-DPACP-162} for more processing details. The input data  are  mostly the time series of field-of-view  transits of the broad band photometry in the calibrated \gmag, \gbp, and \grp bands. Additionally,  for Cepheid and RR~Lyrae stars we use radial velocity time series from the RVS instrument (see section~\ref{ssec:rvsSpectra}) \citep{DR3-DPACP-169, DR3-DPACP-168};  the long-period variable (LPV) analysis included RP~spectral time series \citep[which are not part of \gdr{3},][]{DR3-DPACP-171}; and  short-timescale variables are based on per-CCD \gmag-band photometry. We refer  to \citet{DR3-DPACP-173}  for more information about solar-like variables; \citet{DR3-DPACP-172} for young stellar objects; \citet{DR3-DPACP-166} concerning microlensing events; \citet{DR3-DPACP-174} for ellipsoidal variables with possible compact object secondaries; and  \citet{DR3-DPACP-167} for information on active galactic nuclei (AGN) candidates.
Candidate eclipsing binaries are described in \citet{DR3-DPACP-170}. A subset of those also have parameters in the NSS tables (see section~\ref{ssec:binary}). 

Variability products in \gdrthree can be accessed as follows. In the \textit{gaia\_source} table the field  \textit{phot\_variable\_flag} 
is set to `VARIABLE' when a source appears in any of the \textit{vari\_*} tables, except for the \textit{vari\_summary}\footnote{\varitable{vari\_summary}} table which includes non-variable sources that appear in GAPS (see below). The \textit{vari\_summary} table lists statistical parameters for all variable sources and all sources in GAPS.
The time series in \gmag, \gbp, and \grp bands of all sources listed in the \textit{vari\_summary} table are available  as light curve data through the Datalink interface.  

The quality of this data base is impressive. For instance, concerning LPVs, the catalogue includes about $1.7$~million sources with \gmag variability amplitudes greater than $0.1$~mag ($0.2$~mag in \gdrtwo). The period is derived for about 392\,000 of them. In \gdrtwo we only identified about 151\,000 LPV sources. 
Concerning \gdrthree data, in many cases it was possible to identify the spectral class from RP spectra, that is, cool giants of spectral types M (oxygen-rich) and C (carbon-rich). This classification was based on  the presence in the RP spectra of  numerous molecular absorption bands, mainly those due to TiO and related oxides for M-type and molecular bands mostly associated with CH and C$_2$ molecules for C-type stars \citep{DR3-DPACP-171}.

More than 270\,000 confirmed RR~Lyrae stars are released in \gdrthree,  almost
doubling the \gdrtwo RR~Lyrae catalogue.  In addition, we provide  a better characterisation of the RR~Lyrae pulsational and astrophysical parameters. This, along with the improved astrometry published with \gedrthree, make this sample the largest, most homogeneous all-sky catalogue of  RR~Lyrae stars published so far \citep{DR3-DPACP-168}.  

A small, but significant number of micro-lensing event candidates (\gdrthreevarimulens in total, of which 90 are new) are identified \citep{DR3-DPACP-166}. While testing the exoplanet detection method, two \gaia-discovered  transiting extra-solar planets were found from the epoch photometry \citep{DR3-DPACP-181}. This demonstrates the feasibility of the detection approach and \gaia's potential for discovering exoplanet candidates.  About \gdrthreevariplanets exoplanet candidates  are  released in \gdrthree.

A large number of ellipsoidal
variable candidates were detected. Their  variability is due to the tidal interaction with a companion in a close binary system. About 6\,000 short-period ellipsoidal variables have relatively large-amplitude modulations in \gmag, possibly indicating a  massive, unseen secondary. Among those, 262 systems have a higher probability of having a compact secondary. Follow-up observations are needed to verify the true nature of these variables \citep{DR3-DPACP-174}.

\subsection{\gaia\ Andromeda Photometric Survey}

The photometric time series for all \gaia sources  located within a 5.5$\deg$ radius  centred on the Andromeda galaxy are part of GAPS, which contains more than $1.2$ million sources. Whether or not  a source appears in the GAPS survey is indicated in the \textit{gaia\_source} table by setting the field \textit{in\_andromeda\_survey} 
to `true'. The time-series statistics for GAPS sources are available in the \textit{vari\_summary} table. \citet{DR3-DPACP-142} give more details.

\subsection{Caveats}

As a caveat, a number of sources show more than one type of variability. In general this overlap can be scientifically explained. However, this is not always the case, for instance a few objects were classified as both long period and  short timescale  \citep{DR3-DPACP-171}.  
\cite{DR3-DPACP-164} discuss spurious  periodic variations in the photometric data, due to instrumental effects.

\section{Non-single stars}\label{ssec:binary}

About 800\,000 solutions for NSS including astrometric \citep{DR3-DPACP-163}, spectroscopic  (single lined  SB1; and double lined  SB2) \citep{DR3-DPACP-178}, and eclipsing binaries \citep{DR3-DPACP-179}  are published in \gdrthree, with either orbital elements or trend parameters, or combinations of these.  
\subsection{NSS archive tables}\label{ssec:nssTables}
The NSS tables are organised according to the type of solution: \textit{nss\_two\_body\_orbit}\footnote{\nsstable{nss_two_body_orbit}} contains the orbital parameters for all the binary categories;  \textit{nss\_acceleration\_astro}\footnote{\nsstable{nss_acceleration_astro}} contains accelerations  for sources with an astrometric motion better described using a quadratic or cubic proper motion;  \textit{nss\_non\_linear\_spectro}\footnote{\nsstable{nss_non_linear_spectro}} presents trend (long period) solutions of spectroscopic binaries; \textit{nss\_vim\_fl}\footnote{\nsstable{nss_vim_fl}} includes objects that exhibit photocentre displacements due to the photometric variability of one component, requiring the correction of the astrometric parameters.

This catalogue outnumbers existing surveys by large factors, spanning a large range of binary types, periods, and magnitudes. The potential of the \gdrthree binary star content is outlined in \citet{DR3-DPACP-100}.

\subsection{Caveats}

Binaries can simultaneously belong to different classes; for example, astrometric  binaries can also be  spectroscopic binaries (identified as \textit{astroSpectroSB1}
in the \textit{nss\_solution\_type} in the table \textit{nss\_two\_body\_orbit}) 
or eclipsing binaries can also be spectroscopic binaries (identified as \textit{eclipsingSpectroSB1}).
In many cases of multiple solutions, combined solutions have been computed and included in the \gdrthree catalogue.  
However, combined solutions are not always provided, and  sources can be found  in several tables simultaneously. More information and advice on how to deal with these cases  can be found in \cite{DR3-DPACP-100}.

Acceleration solutions are not always in agreement with expectations from known orbits in external catalogues, and a fraction of them could have had an orbital solution. In general the parallaxes and proper motions derived by the NSS processing are more precise than those derived by the \gaia\ astrometric global iterative solution (see section \ref{ssec:edr3}) which assumes that all stars are single. This is not always the case for the acceleration solutions, which should be used with caution \citep{DR3-DPACP-100, DR3-DPACP-127}.

Spurious solutions around the satellite precession period ($62.97$ days) or for some short periods can be found for SB1. Formal uncertainties are not rescaled according to the goodness of fit for all the binary types, but only for the astrometric solutions. The NSS sample is far from complete. This is because of  a number of selection effects due to data processing and additional filtering. Statistical studies on the data  should take this into account.

\section{Solar System objects}\label{ssec:solar}

In \gdrthree, about 160\,000 SSOs were processed and analysed. As in previous data releases, known SSOs are searched for by matching the observed transits  to computed transits based on  the information on the satellite orbit, the scanning law, and a numerical integration of the SSO motion. \cite{DR3-DPACP-150} gives  a  description of the selection and processing. After filtering the list for possible contaminants, the final input selection had $3\,513\,248$ transits for $156\,837$ known asteroids. Planetary satellites were also added following a  similar procedure. In total, 31 planetary satellites are included. In addition, \gdrthree includes  the astrometry of  unknown moving sources based on the AL motion of objects observed from December 2016 to June 2017. The final input list of unidentified SSOs for \gdrthree  comprises 4522 transits, corresponding to 1531 groups of chained transits of objects that at the time of processing were considered  unmatched. Later, \citet{DR3-DPACP-150} identify  712 SSOs. These sources still appear as unmatched in the table \textit{sso\_source}\footnote{\ssotable{sso_source}}. It cannot be excluded that some of the still unmatched sources can be linked to known objects.

The astrometric accuracy of the orbits is impressive and remains at sub-mas  level for $\gmag<17$, reaching an exceptional value of $\sim0.25$ mas for  $12<$\gmag$<15$ mag.


\gdrthree contains spectro-photometry  for more than $60\,000$ asteroids the majority of which have been observed with \gmag between $\sim 18$ and 20. The internally calibrated \bprp epoch spectra are divided by the solar analogue spectrum to obtain epoch reflectance spectra that are subsequently averaged and sampled in 16 bands in wavelength. Only spectra derived using more than three epochs and with an average S/N of higher than 13 are published. No further rejection is applied. Poor spectra are flagged on a wavelength-by-wavelength basis introducing the \textit{sso\_reflectance\_spectrum\_flag}\footnote{\ssotable{sso_reflectance_spectrum}}, 
an array of 16 integers, one for each wavelength of the spectral bands \citep{DR3-DPACP-89}. The main properties of the reflectance spectra are described in section~\ref{ssec:science}.

\section{Object characterisation} \label{ssec:classif}

\gdrthree includes APs for  stars, galaxies, and QSOs. About 1600 million objects have class probabilities ($G<21$), about 470 million stars have stellar parameters ($G<19$), and there are about 6 million QSOs and  about 1.3 million galaxy candidate redshifts.  More details can be found in \cite{DR3-DPACP-157} concerning the AP content and an overview of the methods used in the software (Apsis) to produce these data. 

\subsection{Data processing}

In total, 13  modules in the Apsis software provide 43 primary APs along with auxiliary parameters in 538  fields which appear in ten  tables  of  the  Gaia archive. A subset of APs is available in \textit{gaia\_source}. The astrophysical characterisation makes use of  \gedr{3} broad band photometry and parallaxes, and \gdr{3} mean RVS spectra and internally calibrated sampled  mean BP/RP spectra.
The stellar APs comprise  atmospheric properties, evolutionary parameters, metallicity, individual chemical element abundances, and extinction parameters, along with other characterisation such as equivalent widths of the H$\alpha$ line and activity index for cool active stars.   

 The discrete source classifier (DSC)  produces the object classification, that is, it assigns class probabilities to all sources for five main classes, using different classifiers; these classes are QSO, galaxy, star, white dwarf, and physical binary star.  The result of this classification is then used by four modules to initiate the processing of galaxies, QSOs, outliers, and extinction. The unresolved  galaxy  classifier (UGC) and the QSO classifier (QSOC) modules \citep{DR3-DPACP-158} provide redshifts for candidate galaxies and QSOs. A more detailed discussion about the \gdrthree extragalactic content can be found in section~\ref{ssec:QSOs}. 

The   OA  module  performs  an  unsupervised classification for sources with lower probabilities from DSC, using SOMs \citep{kohonen_self-organizing_2001}. OA groups similar objects into neurons on a $30\times30$ grid according to the similarity of their \bprp spectra, as reported in \textit{oa\_neuron\_information}\footnote{\linktotableap{oa_neuron_information}} \citep[see][figure 11, for examples]{DR3-DPACP-157}. Other multi-dimensional data comprise a total Galactic extinction map at four healpix levels as well an optimal-level map derived by the total Galactic extinction module (TGE). The above data products are detailed in \cite{DR3-DPACP-158}. 

For the stellar and interstellar medium characterisation, the General Stellar Parametriser for Photometry (\gspphot) derives astrophysical parameters (\teff, \logg, \ag,  distance, \ldots) from \bprp spectra down to $G=19$ assuming they  are all stars \citep{DR3-DPACP-156}. The multiple source classifier (MSC) derives similar stellar parameters under the hypothesis that all objects are unresolved binaries using the same input data as \gspphot. 
The General Stellar Parametriser for Spectroscopy (\gspspec) derives atmospheric parameters using the mean RVS spectra.  It additionally derives 13 chemical species and diffuse interstellar band (DIB) equivalent widths  \citep[see][for details]{DR3-DPACP-186}. 
Other modules analyse only a selected class of objects. This is the case for  the Extended Stellar Parametrisers (ESPs) dealing with emission line stars (ESP-ELS), hot stars ($7000 < \teff<50\,000$~K. ESP-HS), cool stars (ESP-CS),  
and ultra-cool dwarfs ($\teff < 2500$~K, ESP-UCD). These modules together produce  \teff and \logg\ but also $v \sin i$, spectral type classifications, H$\alpha$ equivalent widths, and chromospheric activity index (using the calcium infrared triplet in the RVS domain \citep{DR3-DPACP-175}. 

The Final   Luminosity   Age   Mass   Estimator (FLAME)  derives evolutionary parameters from data that are processed by \gspphot\ and \gspspec.  These comprise the stellar radius, luminosity, gravitational redshift,  mass,  age and evolutionary stage for stars. All of these data products appear in the \textit{astrophysical\_parameters}\footnote{\linktotableap{astrophysical_parameters}} 
and \textit{astrophysical\_parameters\_supp}\footnote{\linktotableap{astrophysical_parameters_supp}} tables in the \gaia\ archive.
In addition to these data products, Markov Chain Monte Carlo  (MCMC) samples are provided for two of the modules, \gspphot  and MSC through the Datalink interface.

To date, this is the most extensive catalogue of astrophysical parameters homogeneously derived. It is based on Gaia-only data, and it will be superseded only by the fourth data release, {\it Gaia} DR4. Until then, it will remain the reference for upcoming ground- and space-based surveys. References to some applications and advice on the use of the APs can be found in the papers listed in section~\ref{ssec:science}.

\subsection{Caveats}
The AP quality, although quite reliable on average, does vary  depending on the quality of the input data (parallaxes, magnitudes, and spectra) but it can also vary based on the assumptions in the methods.  The users should be aware of several issues.
 Variability is not taken into account in the processing, because only mean spectra and mean broad band magnitudes are used as input. Therefore, parameters of stars showing  large variability might be inaccurate.

 APs derived for objects in high-density regions are affected by crowding. For instance [M/H] in the core of a dense cluster can significantly differ from the value in the outskirts.
 
 \gspphot distance estimates are
    underestimated for sources with large parallax errors because of an extinction prior which then dominates the distance inference.
\gspphot metallicity is poorly derived for metal-poor stars ($[\mathrm{M}/\mathrm{H}]<-1$), and a metallicity calibration tool is proposed to the users to calibrate these. The data show residual degeneracy among the parameters, for instance \teff and the extinction in the \gmag, \ag, from \gspphot.

 MSC treats all stars as unresolved binaries in BP/RP spectra; 
 DSC shows poor performance on the classification of physical binaries and white dwarfs.
   
Additional information about the limitations of the APs are detailed in \cite{DR3-DPACP-127}, while both the quality and validation of the stellar and non-stellar content are discussed more extensively in \cite{DR3-DPACP-160} and \cite{DR3-DPACP-158}, respectively. In general, users are advised to use quality flags and follow the recommendations detailed in the above papers.

\section{Extended objects}\label{ssec:extended}

The \gaia on-board system is designed to detect point-like sources \citep{2015AA...576A..74D}. However, EOs such as galaxies and galaxies hosting a QSO can be detected  if their central region is  sufficiently compact and bright. In this case, it is possible to reconstruct a two-dimensional light profile of the extended source. At each passage Gaia observes the nine one-dimensional Astro Field (AF) windows and the two-dimensional Sky Mapper (SM) windows. As the scan angle changes from one observation to the next, after a sufficient number of transits a large part of the source is covered by the observations. Dedicated software can combine the observed windows   on ground to infer the two-dimensional light profile parameters.  
During the first 34 months of operations about $116$ million transits in the AF and SM focal planes were collected and processed. The EO pipeline compares the observations with a large number of simulations of galaxy images deriving the best-fit  surface brightness profile parameters.

The lists of objects that were processed are derived from external catalogues.
The list of QSOs was set up by merging  the  major catalogues of QSOs or AGN candidates published before 2018. In addition, we make use of an unpublished selection of candidates based on \gdrtwo QSOs, and classified on the basis of  their photometric variability \citep{2019A&A...625A..97R}. The list of galaxies was derived from a preliminary analysis  of \gdrtwo sources with a match  in the allWISE catalogue \citep{allWISE}. An unsupervised heuristic method \citep{2022Krone-Martins}  selects  sources for a subsequent analysis.
More than 1 million  previously identified  QSOs are analysed,  identifying a host galaxy around  approximately 64\,000 of them. The surface brightness profiles of the host is published for a subset of about 15\,000 QSOs, with a robust solution.
 Their S\'{e}rsic indexes indicate that they are mostly disc-like galaxies. 
Concerning the galaxy sample, two profiles are published (a free S\'{e}rsic profile and a de Vaucouleurs one).  About $940\,000$ galaxies were analysed and robust solutions were derived for about 914\,000 of them. The distribution of the parameters indicates that Gaia mostly detects elliptical galaxies.

These data can be found in the \gdrthree tables \textit{qso\_candidates}\footnote{\extragal{qso_candidates}} and \textit{galaxy\_candidates}\footnote{\extragal{galaxy_candidates}}. Details on the data processing and input lists can be found in \cite{DR3-DPACP-153}.  This impressive data base is the first all-sky catalogue of  two-dimensional light-profile parameters of  galaxies,  and QSO host galaxies derived at a resolution of about 200~\mas.

 The use of input lists clearly  favours purity over completeness, even if a residual contamination by stars is still present.

\gaia\ observations are limited to  the central area of the galaxy. For this reason, the derived brightness profile parameters do not always agree with external catalogues \citep{DR3-DPACP-127, DR3-DPACP-153}.

\section{\gdrthree extragalactic content}\label{ssec:QSOs}
Extragalactic  objects  are  classified  or  analysed  by  several modules  in  the Gaia data-processing  system using various input data and methods. An overview of the extragalactic processing and content in \gdr{3} is given in \cite{DR3-DPACP-101}.

A predefined list of objects is analysed to derive the surface brightness profiles by the EO module. 
Classification of extragalactic objects  is performed independently by two modules: the VARI module uses photometric light curves, whereas DSC uses the \bprp\ spectra and astrometry. The modules QSOC and UGC estimate the redshifts of QSO and galaxy candidates (respectively) identified by DSC. Finally, the OA module performs a clustering of low-probability classifications from DSC.
 
The fact that different methods, input data, and training data are used to classify or select extragalactic objects has an important consequence: there is no common definition of QSOs or galaxies across the various \gaia modules.
DSC and OA focus on completeness rather than purity.
UGC processes objects with a higher probability of being galaxies according to the DSC classification, whereas QSOC uses a very low threshold on the DSC QSO probability, in order to analyse as many sources as possible. As a result, the sample of QSO candidates with redshifts is complete rather than pure, whereas that of the galaxy candidates with redshifts is of higher purity, although some contamination remains (see below).
 
\subsection{Extragalactic object  tables}
Each one of the above modules provides independent results that are available across several tables. The table \textit{astrophysical\_parameters}
lists all the parameters produced by  DSC. 
The \textit{oa\_neuron\_information}
contains the SOMs from the OA module. The tables \textit{vari\_classifier\_result}\footnote{\varitable{vari_classifier_result}}
and \textit{vari\_agn}\footnote{\varitable{vari_agn}}
present the parameters of the AGNs identified through the photometric light-curve analysis.
From the above tables, we produce two integrated tables in \gdrthree,   \textit{qso\_candidates}
and \textit{galaxy\_candidates},
where all good QSO and galaxy candidates are listed.
The selection rules are detailed in \cite{DR3-DPACP-101}.
The \textit{gaia\_source}
table includes the DSC classification probability as well as two flags, \textit{in\_qso\_candidates}
and \textit{in\_galaxy\_candidates},
which indicate the presence of the source in the integrated table(s).

Already in \gedr{3}, a list of 1.6 million compact extragalactic sources was used to define the \gaia-CRF3 \citep[][]{EDR3-DPACP-133}. These sources were identified from \gedrthree astrometry by means of positional cross-matching with 17 external catalogues of QSOs and AGNs, and were subsequently filtered using astrometric criteria to remove stellar contaminants. The sample spans a magnitude range $13.4<$ \gmag$<21.4$ with a median positional error of $0.45$~\mas. By construction this sample is of high purity. 

\subsection{Completeness and purity}
The integrated  tables of extragalactic objects favour completeness over purity.  
\textit{qso\_candidates} contains 6.6 million candidate QSOs and \textit{galaxy\_candidates} contains 4.8 million candidate galaxies, with global purities estimated to be 50\% and 70\% respectively. Contamination is mostly due to fact that there are many more  stars in the Gaia survey, and arises in part from the choice of probability classification threshold for inclusion. Furthermore, some modules chose not to filter out regions  such as the Magellanic Clouds and  the Galactic plane, which are dense regions of stars and therefore also of contaminants. A very small fraction of sources are too bright to be genuine QSOs or galaxies.  The same astrometric filtering as for \gaia-CRF3 applied to the content of the integrated QSO table \textit{qso\_candidates}\ gives a set of $\sim$ 1.9 million sources, which are identified by the flag
\textit{astrometric\_selection}\
\citep{DR3-DPACP-101}.
Purer subsets of the two tables, containing $\sim $ 1.9 million probable QSOs and $\sim$ 2.9 million probable galaxies (both with around 95\% purity) can be selected using the queries on EO, DSC, and VARI quality flag parameters listed in \citet{DR3-DPACP-101}. Concerning QSOs, this new sample has about $1.8$ million sources in common with the astrometric sample. 

 Finally, the DSC star class could  be used to select a purer star sample, rejecting objects that are  in the  \textit{qso\_candidates} or \textit{galaxy\_candidates}, that is, those identified as candidate galaxies, QSOs, or extended objects. However, concerning extended objects, the DSC has not used morphological information in its classification \citep[see for details][]{DR3-DPACP-158}. Indeed there is no general morphological classification in \gdrthree. Only a limited number of galaxy  and QSO candidates  (based on non-Gaia data)  were analysed for evidence of a host galaxy using 2D morphology built up from the 1D scans (see section~\ref{ssec:extended}). As a consequence,  residual  contamination from extragalactic sources can still be present in a star sample selected following the above criteria.

\section{The scientific performances of the catalogue}\label{ssec:science}

The scientific quality and the potential of the \gdrthree data are demonstrated in a number of accompanying papers providing basic science applications and additional validation of the catalogue. These illustrate the limitations of the data, providing advice and guidance to the users on specific scientific problems. The following topics are treated. 

\citet{DR3-DPACP-123} present clean samples of very high-quality   astrophysical parameters of stars  derived from low-resolution \bprp and  RVS spectra all across the HR diagram,  selected through severe quality cuts.  These data have a long-term legacy value for future follow-up studies and missions and are released together with \gdrthree in separate tables. These samples include about 3 million OBA young disc stars, roughly 3 million FGKM stars, and  about 21\,000  ultra-cool dwarfs. \citet{DR3-DPACP-123}  identify specific subsamples  of particular interest to the community such as approximately  6\,000 solar-analogues and 15\,000 carbon stars, and provide homogeneous parameters for a subsample of the $111$ \gaia spectro-photometric standard stars  defined by \citet{DR3-DPACP-137}.   For the ultra-cool dwarfs \gaia\ and 2MASS data are combined to provide radius and luminosity. 

\citet{DR3-DPACP-104}  use \gaia\ astrometry, radial velocities, chemo-dynamical analysis of disc and halo populations, producing the largest  all-sky chemical map to date and analysing the abundances of streams from accretion events. This map includes chemical information for 597 open clusters, which represents the largest compilation of abundances homogeneously derived for clusters.

\citet{DR3-DPACP-144} discuss the distribution of the DIB at 862 nm in the RVS spectral range in connection with the interstellar extinction to within a few kiloparsecs of the Sun. The paper provides the most precise  measurement to date of the rest frame wavelength of the DIB at  $8620.86\pm 0.019$~\ang. 

\citet{DR3-DPACP-75} explore  non-axisymmetric features in the disc of the Milky Way   in both configuration and velocity space. These authors use \gdrthree APs  and variability classifications to select various stellar populations in the disc of the Milky Way, tracing the spiral structure up to 4--5~\kpc from the Sun. About 6 million red giant branch stars  allow the authors to map the velocity field in the disc up to 8~\kpc from the Sun, allowing the detection of signatures of the Galactic bar. 

\citet{DR3-DPACP-100} present a clean catalogue of binary stars,  discussing its completeness and some  statistical features of the orbital elements  in comparison with external catalogues.  In addition, a catalogue of tens of thousands of stellar masses is provided. Several compact object candidates are identified.
The catalogue includes sources found  in  rare evolutionary stages such as EL CVn,  underlining the potential of analysing both photometric and orbital data.
 New binary UCD  candidates are discovered and their masses estimated; two new exo-planets are found, and several dozen  candidates are identified.
The catalogue is made available together with \gdrthree tables.

\citet{DR3-DPACP-89} present reflectance spectra, derived from \bprp spectra, for SSOs, discussing their  statistical properties. The $(z-i)$ colours are derived from the spectra, where $z$ and $i$ represent the values of the reflectances estimated with spline interpolation at 748 and 893~nm, respectively. $(z-i)$ allows the identification of asteroid families sharing similar properties and common origin. An interesting feature is that  the spectral slope\footnote{the spectral slope is derived by fitting the mean reflectance spectra in the wavelength range 450 and 760~nm  with a straight line and taking the angular coefficient} of the mean reflectance spectra and the depth of the 1~$\mu$m absorption band seem to show an increase with time for the S-type family, which is possibly due to space weathering effects on the asteroid surfaces. The catalogue is included in the data release.

\citet{DR3-DPACP-93} demonstrate the potential of synthetic photometry obtained  from flux-calibrated \bprp spectra for  pass-bands fully enclosed in the \gaia\ wavelength range. This photometry has been used internally for validation purposes. Synthetic photometry in various photometric systems can be produced using the Python package GaiaXPy (see section~\ref{ssec:tools}).  High-quality external photometry in large and medium passbands is reproduced at a few percent level in general and up to the milli-mag  level when the synthetic photometry is standardised using an external reference catalogue. For a subset of 13 wide and medium bands, we release the  Gaia Synthetic Photometry Catalogue (GSPC), an all-sky space-based catalogue of standardised photometry for the majority of the stars with released spectra and \gmag$<17.65$.  A separate catalogue contains synthetic photometry in a selection of relevant bands for 100\,000 white dwarfs selected from  \gedrthree.  

Lastly, \citet{DR3-DPACP-79} investigate the properties of high-mass main sequence pulsators, showing that \gdrthree data are precise and accurate enough to identify   nearby OBAF-type pulsators.

\section{Software tools}\label{ssec:tools}
A number of Python software tools\footnote{software tools and the related documentation are available at {\url{https://www.cosmos.esa.int/web/gaia/dr3-software-tools}}} are made available to the scientific community.  The following are some examples.

\begin{itemize}
    \item The Python package GaiaXPy which offers  the possibility to compute  synthetic photometry from \bprp spectra in various photometric systems \citep[see][and Sect.~\ref{ssec:science}]{DR3-DPACP-93}, and to simulate \bprp mean spectra from a custom input absolute spectral energy distribution. In addition sampled mean spectra can be derived from the continuous basis function representation. Detailed information can be found in  \citet{EDR3-DPACP-118}.
\item A tool to  re-calibrate the \gspphot metallicity output  \citep[see][Appendix E for detail]{DR3-DPACP-157} based on comparison with external surveys.
    \item A package  to re-calibrate the \gspspec metallicity, $[\alpha/\mathrm{Fe}]$, and \loggrav \citep{DR3-DPACP-186}.
    \item  A script to derive bolometric corrections for the stars analysed by FLAME.
    \item A package to extract the extinction as a function of Galactic coordinates from the two-dimensional TGE Galactic maps.
    \item A tool to visualise the neurons in the OA SOMs.
    \item A tutorial on the programmatic download of large sets of Datalink items ($>5000$ elements).
    \item Examples of cone search scripts.
\end{itemize}

 We point out that the  \gspphot and \gspspec re-calibration tools proposed here have their own limitations, as described in the related documentation.  They provide  tentative expressions to be used at the  discretion of the user, depending on the science case.

\section{Conclusions}\label{ssec:conclusions}
 
\begin{figure*}
    \sidecaption
    \includegraphics[width=12cm]{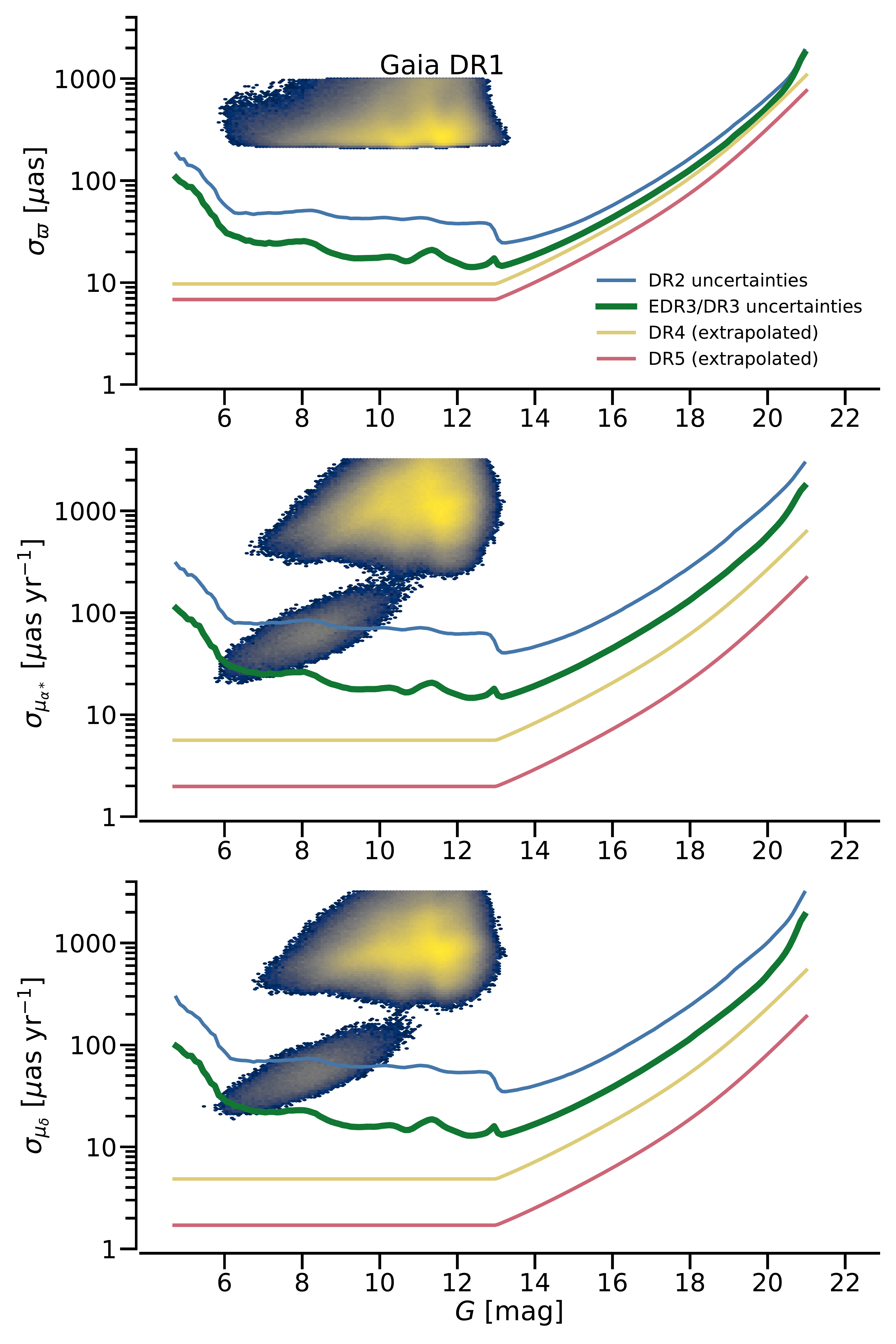}
    \caption{Uncertainties on the astrometric parameters vs.\ $G$ for \gaia\ data releases 1--3 and for the future releases \gdr{4} and \gdr{5}. The panels show from top to bottom the uncertainties in parallax, proper motion in Right Ascension, and proper motion in Declination. The uncertainties for \gdr{1} refer to the Tycho-\gaia\ Astrometric Solution and are shown in the form of density maps, with lighter colours indicating a higher density of sources. The two distinct low-uncertainty elliptical regions in the \gdr{1} proper motion uncertainties are due to stars for which the \textit{Hipparcos} and \gaia\ positions could be combined to derive proper motions over a 24-year time baseline. The proper motion uncertainties for \gdr{3}, based on only a 34~month time baseline are comparable or even slightly better.}
    \label{fig:drperformance}
\end{figure*}

\gdrthree expands \gedrthree with a rich and large set of data products containing detailed astrophysical information for the same source catalogue. The quantity, quality and variety of astrophysical data constitute a major advance in the series of \gaia\ data releases. This release includes the largest collection of  all-sky spectrophotometry and radial velocities and the largest collection of  variable sources ever produced. The availability of APs from low- and high-resolution spectra greatly surpasses existing catalogues. This catalogue includes a spectrophotometric and dynamical survey of SSOs of the highest accuracy. The NSS  content outnumbers all existing binary star catalogues, and \gdr{3} also contains the first all-sky space-based survey of QSO host and galaxy two-dimensional light profiles.

In this paper we briefly summarise the main additions  and improvements to the data processing and we include comments on the data quality. However, the \gdrthree data products are numerous and complex and not all the details could be presented here. A more complete overview, insights into known issues with the data, and advice on the use of the data can be found in  the papers accompanying the release and in the online documentation.

The nominal \gaia\ mission ended on July 2019. The  mission  has been extended since then to the end of 2022, with an indicative approval to extend the mission to the end of 2025. In 2025, the propellant for the micro-propulsion system is expected to be exhausted, and the precision on the attitude and spin rate of the satellite required for the astrometry can then no longer be maintained. At this point, the \gaia\ mission end of life will be reached and  over 10 years of data will have been collected. In this context, two further data releases are foreseen, \gdr{4} and \gdr{5} which will include the data from the extended mission. \gdr{4} will be based on 66 months of data, including a six-month period when the satellite was operated with a reversed direction of the precession of the spin axis around the direction to the Sun. This will mitigate the degeneracy between the across scan motion of the sources and their parallaxes, reducing this specific source of systematic errors on the astrometry. A major new aspect of \gdr{4} is that it will include all time series data, that is, epoch astrometry, broad band photometry, radial velocities, and epoch BP, RP, and RVS spectra for all sources.  In addition,   we plan to release full astrometric, photometric, and radial-velocity catalogues; all the available variable-star and NSS solutions; an extended sample of  source classifications and multiple APs for stars, and extragalactic objects. Finally, an updated extra-solar planet list will be published. The publication of \gdr{4} is expected not before the end of 2025. \gdr{5} will be the final release from the {\gaia} mission, and will be based on data collected over the full nominal plus extended mission periods and including all the data products mentioned above, as well as the {\gaia} legacy archive. \gdr{5} is expected not before the end of 2030. 

\figrefalt{fig:drperformance} summarises the astrometric uncertainties of the \gaia\ data releases so far and also shows the extrapolated uncertainties for \gdr{4} and \gdr{5}. The latter are extrapolated from the \gdr{3} performance according to the amount of data collected, $5.5$~years and an expected $10$~years, respectively\footnote{\url{https://www.cosmos.esa.int/web/gaia/science-performance}}. With time the signal-to-noise ratio for all \gaia\ data products, including parallaxes, improves as $\sqrt{t}$, and therefore the precisions for \gdr{4} and \gdr{5} are expected to improve by  factors of $1.4$ and $1.9$, respectively. For the proper motions, the improvement goes as $t^{1.5}$, which means improvements by factors of $2.7$ and $6.6$ with respect to \gdr{3}. For a more extensive discussion of the expected gains in future \gaia\ data releases, we refer to \cite{2021ARA&A..59...59B}.

There is therefore still much more to look forward to, but for now we invite the reader to explore the veritable supermarket of astronomical and astrophysical information that is \gdr{3}.

\section*{Acknowledgements\label{sec:acknowl}}
\addcontentsline{toc}{chapter}{Acknowledgements}
This work presents results from the European Space Agency (ESA) space mission \gaia. \gaia\ data are being processed by the \gaia\ Data Processing and Analysis Consortium (DPAC). Funding for the DPAC is provided by national institutions, in particular the institutions participating in the \gaia\ MultiLateral Agreement (MLA). The \gaia\ mission website is \url{https://www.cosmos.esa.int/gaia}. The \gaia\ archive website is \url{https://archives.esac.esa.int/gaia}.
Acknowledgements are given in Appendix~\ref{ssec:appendixA}

\bibliographystyle{aa}
\bibliography{AA43940-22}
\begin{appendix}
\section{}\label{ssec:appendixA}
The \gaia\ mission and data processing have financially been supported by, in alphabetical order by country:
\begin{itemize}
\item the Algerian Centre de Recherche en Astronomie, Astrophysique et G\'{e}ophysique of Bouzareah Observatory;
\item the Austrian Fonds zur F\"{o}rderung der wissenschaftlichen Forschung (FWF) Hertha Firnberg Programme through grants T359, P20046, and P23737;
\item the BELgian federal Science Policy Office (BELSPO) through various PROgramme de D\'{e}veloppement d'Exp\'{e}riences scientifiques (PRODEX) grants and the Polish Academy of Sciences - Fonds Wetenschappelijk Onderzoek through grant VS.091.16N, and the Fonds de la Recherche Scientifique (FNRS), and the Research Council of Katholieke Universiteit (KU) Leuven through grant C16/18/005 (Pushing AsteRoseismology to the next level with TESS, GaiA, and the Sloan DIgital Sky SurvEy -- PARADISE);  
\item the Brazil-France exchange programmes Funda\c{c}\~{a}o de Amparo \`{a} Pesquisa do Estado de S\~{a}o Paulo (FAPESP) and Coordena\c{c}\~{a}o de Aperfeicoamento de Pessoal de N\'{\i}vel Superior (CAPES) - Comit\'{e} Fran\c{c}ais d'Evaluation de la Coop\'{e}ration Universitaire et Scientifique avec le Br\'{e}sil (COFECUB);
\item the Chilean Agencia Nacional de Investigaci\'{o}n y Desarrollo (ANID) through Fondo Nacional de Desarrollo Cient\'{\i}fico y Tecnol\'{o}gico (FONDECYT) Regular Project 1210992 (L.~Chemin);
\item the National Natural Science Foundation of China (NSFC) through grants 11573054, 11703065, and 12173069, the China Scholarship Council through grant 201806040200, and the Natural Science Foundation of Shanghai through grant 21ZR1474100;  
\item the Tenure Track Pilot Programme of the Croatian Science Foundation and the \'{E}cole Polytechnique F\'{e}d\'{e}rale de Lausanne and the project TTP-2018-07-1171 `Mining the Variable Sky', with the funds of the Croatian-Swiss Research Programme;
\item the Czech-Republic Ministry of Education, Youth, and Sports through grant LG 15010 and INTER-EXCELLENCE grant LTAUSA18093, and the Czech Space Office through ESA PECS contract 98058;
\item the Danish Ministry of Science;
\item the Estonian Ministry of Education and Research through grant IUT40-1;
\item the European Commission’s Sixth Framework Programme through the European Leadership in Space Astrometry (\href{https://www.cosmos.esa.int/web/gaia/elsa-rtn-programme}{ELSA}) Marie Curie Research Training Network (MRTN-CT-2006-033481), through Marie Curie project PIOF-GA-2009-255267 (Space AsteroSeismology \& RR Lyrae stars, SAS-RRL), and through a Marie Curie Transfer-of-Knowledge (ToK) fellowship (MTKD-CT-2004-014188); the European Commission's Seventh Framework Programme through grant FP7-606740 (FP7-SPACE-2013-1) for the \gaia\ European Network for Improved data User Services (\href{https://gaia.ub.edu/twiki/do/view/GENIUS/}{GENIUS}) and through grant 264895 for the \gaia\ Research for European Astronomy Training (\href{https://www.cosmos.esa.int/web/gaia/great-programme}{GREAT-ITN}) network;
\item the European Cooperation in Science and Technology (COST) through COST Action CA18104 `Revealing the Milky Way with \gaia (MW-Gaia)';
\item the European Research Council (ERC) through grants 320360, 647208, and 834148 and through the European Union’s Horizon 2020 research and innovation and excellent science programmes through Marie Sk{\l}odowska-Curie grant 745617 (Our Galaxy at full HD -- Gal-HD) and 895174 (The build-up and fate of self-gravitating systems in the Universe) as well as grants 687378 (Small Bodies: Near and Far), 682115 (Using the Magellanic Clouds to Understand the Interaction of Galaxies), 695099 (A sub-percent distance scale from binaries and Cepheids -- CepBin), 716155 (Structured ACCREtion Disks -- SACCRED), 951549 (Sub-percent calibration of the extragalactic distance scale in the era of big surveys -- UniverScale), and 101004214 (Innovative Scientific Data Exploration and Exploitation Applications for Space Sciences -- EXPLORE);
\item the European Science Foundation (ESF), in the framework of the \gaia\ Research for European Astronomy Training Research Network Programme (\href{https://www.cosmos.esa.int/web/gaia/great-programme}{GREAT-ESF});
\item the European Space Agency (ESA) in the framework of the \gaia\ project, through the Plan for European Cooperating States (PECS) programme through contracts C98090 and 4000106398/12/NL/KML for Hungary, through contract 4000115263/15/NL/IB for Germany, and through PROgramme de D\'{e}veloppement d'Exp\'{e}riences scientifiques (PRODEX) grant 4000127986 for Slovenia;  
\item the Academy of Finland through grants 299543, 307157, 325805, 328654, 336546, and 345115 and the Magnus Ehrnrooth Foundation;
\item the French Centre National d’\'{E}tudes Spatiales (CNES), the Agence Nationale de la Recherche (ANR) through grant ANR-10-IDEX-0001-02 for the `Investissements d'avenir' programme, through grant ANR-15-CE31-0007 for project `Modelling the Milky Way in the \gaia era’ (MOD4Gaia), through grant ANR-14-CE33-0014-01 for project `The Milky Way disc formation in the \gaia era’ (ARCHEOGAL), through grant ANR-15-CE31-0012-01 for project `Unlocking the potential of Cepheids as primary distance calibrators’ (UnlockCepheids), through grant ANR-19-CE31-0017 for project `Secular evolution of galxies' (SEGAL), and through grant ANR-18-CE31-0006 for project `Galactic Dark Matter' (GaDaMa), the Centre National de la Recherche Scientifique (CNRS) and its SNO \gaia of the Institut des Sciences de l’Univers (INSU), its Programmes Nationaux: Cosmologie et Galaxies (PNCG), Gravitation R\'{e}f\'{e}rences Astronomie M\'{e}trologie (PNGRAM), Plan\'{e}tologie (PNP), Physique et Chimie du Milieu Interstellaire (PCMI), and Physique Stellaire (PNPS), the `Action F\'{e}d\'{e}ratrice \gaia' of the Observatoire de Paris, the R\'{e}gion de Franche-Comt\'{e}, the Institut National Polytechnique (INP) and the Institut National de Physique nucl\'{e}aire et de Physique des Particules (IN2P3) co-funded by CNES;
\item the German Aerospace Agency (Deutsches Zentrum f\"{u}r Luft- und Raumfahrt e.V., DLR) through grants 50QG0501, 50QG0601, 50QG0602, 50QG0701, 50QG0901, 50QG1001, 50QG1101, 50\-QG1401, 50QG1402, 50QG1403, 50QG1404, 50QG1904, 50QG2101, 50QG2102, and 50QG2202, and the Centre for Information Services and High Performance Computing (ZIH) at the Technische Universit\"{a}t Dresden for generous allocations of computer time;
\item the Hungarian Academy of Sciences through the Lend\"{u}let Programme grants LP2014-17 and LP2018-7 and the Hungarian National Research, Development, and Innovation Office (NKFIH) through grant KKP-137523 (`SeismoLab');
\item the Science Foundation Ireland (SFI) through a Royal Society - SFI University Research Fellowship (M.~Fraser);
\item the Israel Ministry of Science and Technology through grant 3-18143 and the Tel Aviv University Center for Artificial Intelligence and Data Science (TAD) through a grant;
\item the Agenzia Spaziale Italiana (ASI) through contracts I/037/08/0, I/058/10/0, 2014-025-R.0, 2014-025-R.1.2015, and 2018-24-HH.0 to the Italian Istituto Nazionale di Astrofisica (INAF), contract 2014-049-R.0/1/2 to INAF for the Space Science Data Centre (SSDC, formerly known as the ASI Science Data Center, ASDC), contracts I/008/10/0, 2013/030/I.0, 2013-030-I.0.1-2015, and 2016-17-I.0 to the Aerospace Logistics Technology Engineering Company (ALTEC S.p.A.), INAF, and the Italian Ministry of Education, University, and Research (Ministero dell'Istruzione, dell'Universit\`{a} e della Ricerca) through the Premiale project `MIning The Cosmos Big Data and Innovative Italian Technology for Frontier Astrophysics and Cosmology' (MITiC);
\item the Netherlands Organisation for Scientific Research (NWO) through grant NWO-M-614.061.414, through a VICI grant (A.~Helmi), and through a Spinoza prize (A.~Helmi), and the Netherlands Research School for Astronomy (NOVA);
\item the Polish National Science Centre through HARMONIA grant 2018/30/M/ST9/00311 and DAINA grant 2017/27/L/ST9/03221 and the Ministry of Science and Higher Education (MNiSW) through grant DIR/WK/2018/12;
\item the Portuguese Funda\c{c}\~{a}o para a Ci\^{e}ncia e a Tecnologia (FCT) through national funds, grants SFRH/\-BD/128840/2017 and PTDC/FIS-AST/30389/2017, and work contract DL 57/2016/CP1364/CT0006, the Fundo Europeu de Desenvolvimento Regional (FEDER) through grant POCI-01-0145-FEDER-030389 and its Programa Operacional Competitividade e Internacionaliza\c{c}\~{a}o (COMPETE2020) through grants UIDB/04434/2020 and UIDP/04434/2020, and the Strategic Programme UIDB/\-00099/2020 for the Centro de Astrof\'{\i}sica e Gravita\c{c}\~{a}o (CENTRA);  
\item the Slovenian Research Agency through grant P1-0188;
\item the Spanish Ministry of Economy (MINECO/FEDER, UE), the Spanish Ministry of Science and Innovation (MICIN), the Spanish Ministry of Education, Culture, and Sports, and the Spanish Government through grants BES-2016-078499, BES-2017-083126, BES-C-2017-0085, ESP2016-80079-C2-1-R, ESP2016-80079-C2-2-R, FPU16/03827, PDC2021-121059-C22, RTI2018-095076-B-C22, and TIN2015-65316-P (`Computaci\'{o}n de Altas Prestaciones VII'), the Juan de la Cierva Incorporaci\'{o}n Programme (FJCI-2015-2671 and IJC2019-04862-I for F.~Anders), the Severo Ochoa Centre of Excellence Programme (SEV2015-0493), and MICIN/AEI/10.13039/501100011033 (and the European Union through European Regional Development Fund `A way of making Europe') through grant RTI2018-095076-B-C21, the Institute of Cosmos Sciences University of Barcelona (ICCUB, Unidad de Excelencia `Mar\'{\i}a de Maeztu’) through grant CEX2019-000918-M, the University of Barcelona's official doctoral programme for the development of an R+D+i project through an Ajuts de Personal Investigador en Formaci\'{o} (APIF) grant, the Spanish Virtual Observatory through project AyA2017-84089, the Galician Regional Government, Xunta de Galicia, through grants ED431B-2021/36, ED481A-2019/155, and ED481A-2021/296, the Centro de Investigaci\'{o}n en Tecnolog\'{\i}as de la Informaci\'{o}n y las Comunicaciones (CITIC), funded by the Xunta de Galicia and the European Union (European Regional Development Fund -- Galicia 2014-2020 Programme), through grant ED431G-2019/01, the Red Espa\~{n}ola de Supercomputaci\'{o}n (RES) computer resources at MareNostrum, the Barcelona Supercomputing Centre - Centro Nacional de Supercomputaci\'{o}n (BSC-CNS) through activities AECT-2017-2-0002, AECT-2017-3-0006, AECT-2018-1-0017, AECT-2018-2-0013, AECT-2018-3-0011, AECT-2019-1-0010, AECT-2019-2-0014, AECT-2019-3-0003, AECT-2020-1-0004, and DATA-2020-1-0010, the Departament d'Innovaci\'{o}, Universitats i Empresa de la Generalitat de Catalunya through grant 2014-SGR-1051 for project `Models de Programaci\'{o} i Entorns d'Execuci\'{o} Parallels' (MPEXPAR), and Ramon y Cajal Fellowship RYC2018-025968-I funded by MICIN/AEI/10.13039/501100011033 and the European Science Foundation (`Investing in your future');
\item the Swedish National Space Agency (SNSA/Rymdstyrelsen);
\item the Swiss State Secretariat for Education, Research, and Innovation through the Swiss Activit\'{e}s Nationales Compl\'{e}mentaires and the Swiss National Science Foundation through an Eccellenza Professorial Fellowship (award PCEFP2\_194638 for R.~Anderson);
\item the United Kingdom Particle Physics and Astronomy Research Council (PPARC), the United Kingdom Science and Technology Facilities Council (STFC), and the United Kingdom Space Agency (UKSA) through the following grants to the University of Bristol, the University of Cambridge, the University of Edinburgh, the University of Leicester, the Mullard Space Sciences Laboratory of University College London, and the United Kingdom Rutherford Appleton Laboratory (RAL): PP/D006511/1, PP/D006546/1, PP/D006570/1, ST/I000852/1, ST/J005045/1, ST/K00056X/1, ST/\-K000209/1, ST/K000756/1, ST/L006561/1, ST/N000595/1, ST/N000641/1, ST/N000978/1, ST/\-N001117/1, ST/S000089/1, ST/S000976/1, ST/S000984/1, ST/S001123/1, ST/S001948/1, ST/\-S001980/1, ST/S002103/1, ST/V000969/1, ST/W002469/1, ST/W002493/1, ST/W002671/1, ST/W002809/1, and EP/V520342/1.
\end{itemize}

The \gaia\ project and data processing have made use of:
\begin{itemize}
\item the Set of Identifications, Measurements, and Bibliography for Astronomical Data \citep[SIMBAD,][]{2000AAS..143....9W}, the `Aladin sky atlas' \citep{2000A&AS..143...33B,2014ASPC..485..277B}, and the VizieR catalogue access tool \citep{2000A&AS..143...23O}, all operated at the Centre de Donn\'{e}es astronomiques de Strasbourg (\href{http://cds.u-strasbg.fr/}{CDS});
\item the National Aeronautics and Space Administration (NASA) Astrophysics Data System (\href{http://adsabs.harvard.edu/abstract_service.html}{ADS});
\item the SPace ENVironment Information System (SPENVIS), initiated by the Space Environment and Effects Section (TEC-EES) of ESA and developed by the Belgian Institute for Space Aeronomy (BIRA-IASB) under ESA contract through ESA’s General Support Technologies Programme (GSTP), administered by the BELgian federal Science Policy Office (BELSPO);
\item the software products \href{http://www.starlink.ac.uk/topcat/}{TOPCAT}, \href{http://www.starlink.ac.uk/stil}{STIL}, and \href{http://www.starlink.ac.uk/stilts}{STILTS} \citep{2005ASPC..347...29T,2006ASPC..351..666T};
\item Matplotlib \citep{Hunter:2007};
\item IPython \citep{PER-GRA:2007};  
\item Astropy, a community-developed core Python package for Astronomy \citep{2018AJ....156..123A};
\item R \citep{RManual};
\item Vaex \citep{2018A&A...618A..13B};
\item the \hip-2\ catalogue \citep{2007A&A...474..653V}. The \hip\ and \tyc\ catalogues were constructed under the responsibility of large scientific teams collaborating with ESA. The Consortia Leaders were Lennart Lindegren (Lund, Sweden: NDAC) and Jean Kovalevsky (Grasse, France: FAST), together responsible for the \hip\ Catalogue; Erik H{\o}g (Copenhagen, Denmark: TDAC) responsible for the \tyc\ Catalogue; and Catherine Turon (Meudon, France: INCA) responsible for the \hip\ Input Catalogue (HIC);  
\item the \tyctwo\ catalogue \citep{2000A&A...355L..27H}, the construction of which was supported by the Velux Foundation of 1981 and the Danish Space Board;
\item The \tyc\ double star catalogue \citep[TDSC,][]{2002A&A...384..180F}, based on observations made with the ESA \hip\ astrometry satellite, as supported by the Danish Space Board and the United States Naval Observatory through their double-star programme;
\item data products from the Two Micron All Sky Survey \citep[2MASS,][]{2006AJ....131.1163S}, which is a joint project of the University of Massachusetts and the Infrared Processing and Analysis Center (IPAC) / California Institute of Technology, funded by the National Aeronautics and Space Administration (NASA) and the National Science Foundation (NSF) of the USA;
\item the ninth data release of the AAVSO Photometric All-Sky Survey (\href{https://www.aavso.org/apass}{APASS}, \citealt{apass9}), funded by the Robert Martin Ayers Sciences Fund;
\item the first data release of the Pan-STARRS survey \citep{panstarrs1,panstarrs1b,panstarrs1c,panstarrs1d,panstarrs1e,panstarrs1f}. The Pan-STARRS1 Surveys (PS1) and the PS1 public science archive have been made possible through contributions by the Institute for Astronomy, the University of Hawaii, the Pan-STARRS Project Office, the Max-Planck Society and its participating institutes, the Max Planck Institute for Astronomy, Heidelberg and the Max Planck Institute for Extraterrestrial Physics, Garching, The Johns Hopkins University, Durham University, the University of Edinburgh, the Queen's University Belfast, the Harvard-Smithsonian Center for Astrophysics, the Las Cumbres Observatory Global Telescope Network Incorporated, the National Central University of Taiwan, the Space Telescope Science Institute, the National Aeronautics and Space Administration (NASA) through grant NNX08AR22G issued through the Planetary Science Division of the NASA Science Mission Directorate, the National Science Foundation through grant AST-1238877, the University of Maryland, Eotvos Lorand University (ELTE), the Los Alamos National Laboratory, and the Gordon and Betty Moore Foundation;
\item the second release of the Guide Star Catalogue \citep[GSC2.3,][]{2008AJ....136..735L}. The Guide Star Catalogue II is a joint project of the Space Telescope Science Institute (STScI) and the Osservatorio Astrofisico di Torino (OATo). STScI is operated by the Association of Universities for Research in Astronomy (AURA), for the National Aeronautics and Space Administration (NASA) under contract NAS5-26555. OATo is operated by the Italian National Institute for Astrophysics (INAF). Additional support was provided by the European Southern Observatory (ESO), the Space Telescope European Coordinating Facility (STECF), the International GEMINI project, and the European Space Agency (ESA) Astrophysics Division (nowadays SCI-S);
\item the eXtended, Large (XL) version of the catalogue of Positions and Proper Motions \citep[PPM-XL,][]{2010AJ....139.2440R};
\item data products from the Wide-field Infrared Survey Explorer (WISE), which is a joint project of the University of California, Los Angeles, and the Jet Propulsion Laboratory/California Institute of Technology, and NEOWISE, which is a project of the Jet Propulsion Laboratory/California Institute of Technology. WISE and NEOWISE are funded by the National Aeronautics and Space Administration (NASA);
\item the first data release of the United States Naval Observatory (USNO) Robotic Astrometric Telescope \citep[URAT-1,][]{urat1};
\item the fourth data release of the United States Naval Observatory (USNO) CCD Astrograph Catalogue \citep[UCAC-4,][]{2013AJ....145...44Z};
\item the sixth and final data release of the Radial Velocity Experiment \citep[RAVE DR6,][]{2020AJ....160...83S,rave6a}. Funding for RAVE has been provided by the Leibniz Institute for Astrophysics Potsdam (AIP), the Australian Astronomical Observatory, the Australian National University, the Australian Research Council, the French National Research Agency, the German Research Foundation (SPP 1177 and SFB 881), the European Research Council (ERC-StG 240271 Galactica), the Istituto Nazionale di Astrofisica at Padova, the Johns Hopkins University, the National Science Foundation of the USA (AST-0908326), the W.M.\ Keck foundation, the Macquarie University, the Netherlands Research School for Astronomy, the Natural Sciences and Engineering Research Council of Canada, the Slovenian Research Agency, the Swiss National Science Foundation, the Science \& Technology Facilities Council of the UK, Opticon, Strasbourg Observatory, and the Universities of Basel, Groningen, Heidelberg, and Sydney. The RAVE website is at \url{https://www.rave-survey.org/};
\item the first data release of the Large sky Area Multi-Object Fibre Spectroscopic Telescope \citep[LAMOST DR1,][]{LamostDR1};
\item the K2 Ecliptic Plane Input Catalogue \citep[EPIC,][]{epic-2016ApJS..224....2H};
\item the ninth data release of the Sloan Digitial Sky Survey \citep[SDSS DR9,][]{SDSS9}. Funding for SDSS-III has been provided by the Alfred P. Sloan Foundation, the Participating Institutions, the National Science Foundation, and the United States Department of Energy Office of Science. The SDSS-III website is \url{http://www.sdss3.org/}. SDSS-III is managed by the Astrophysical Research Consortium for the Participating Institutions of the SDSS-III Collaboration including the University of Arizona, the Brazilian Participation Group, Brookhaven National Laboratory, Carnegie Mellon University, University of Florida, the French Participation Group, the German Participation Group, Harvard University, the Instituto de Astrof\'{\i}sica de Canarias, the Michigan State/Notre Dame/JINA Participation Group, Johns Hopkins University, Lawrence Berkeley National Laboratory, Max Planck Institute for Astrophysics, Max Planck Institute for Extraterrestrial Physics, New Mexico State University, New York University, Ohio State University, Pennsylvania State University, University of Portsmouth, Princeton University, the Spanish Participation Group, University of Tokyo, University of Utah, Vanderbilt University, University of Virginia, University of Washington, and Yale University;
\item the thirteenth release of the Sloan Digital Sky Survey \citep[SDSS DR13,][]{2017ApJS..233...25A}. Funding for SDSS-IV has been provided by the Alfred P. Sloan Foundation, the United States Department of Energy Office of Science, and the Participating Institutions. SDSS-IV acknowledges support and resources from the Center for High-Performance Computing at the University of Utah. The SDSS web site is \url{https://www.sdss.org/}. SDSS-IV is managed by the Astrophysical Research Consortium for the Participating Institutions of the SDSS Collaboration including the Brazilian Participation Group, the Carnegie Institution for Science, Carnegie Mellon University, the Chilean Participation Group, the French Participation Group, Harvard-Smithsonian Center for Astrophysics, Instituto de Astrof\'isica de Canarias, The Johns Hopkins University, Kavli Institute for the Physics and Mathematics of the Universe (IPMU) / University of Tokyo, the Korean Participation Group, Lawrence Berkeley National Laboratory, Leibniz Institut f\"ur Astrophysik Potsdam (AIP),  Max-Planck-Institut f\"ur Astronomie (MPIA Heidelberg), Max-Planck-Institut f\"ur Astrophysik (MPA Garching), Max-Planck-Institut f\"ur Extraterrestrische Physik (MPE), National Astronomical Observatories of China, New Mexico State University, New York University, University of Notre Dame, Observat\'ario Nacional / MCTI, The Ohio State University, Pennsylvania State University, Shanghai Astronomical Observatory, United Kingdom Participation Group, Universidad Nacional Aut\'onoma de M\'{e}xico, University of Arizona, University of Colorado Boulder, University of Oxford, University of Portsmouth, University of Utah, University of Virginia, University of Washington, University of Wisconsin, Vanderbilt University, and Yale University;
\item the second release of the SkyMapper catalogue \citep[SkyMapper DR2,][Digital Object Identifier 10.25914/5ce60d31ce759]{2019PASA...36...33O}. The national facility capability for SkyMapper has been funded through grant LE130100104 from the Australian Research Council (ARC) Linkage Infrastructure, Equipment, and Facilities (LIEF) programme, awarded to the University of Sydney, the Australian National University, Swinburne University of Technology, the University of Queensland, the University of Western Australia, the University of Melbourne, Curtin University of Technology, Monash University, and the Australian Astronomical Observatory. SkyMapper is owned and operated by The Australian National University's Research School of Astronomy and Astrophysics. The survey data were processed and provided by the SkyMapper Team at the the Australian National University. The SkyMapper node of the All-Sky Virtual Observatory (ASVO) is hosted at the National Computational Infrastructure (NCI). Development and support the SkyMapper node of the ASVO has been funded in part by Astronomy Australia Limited (AAL) and the Australian Government through the Commonwealth's Education Investment Fund (EIF) and National Collaborative Research Infrastructure Strategy (NCRIS), particularly the National eResearch Collaboration Tools and Resources (NeCTAR) and the Australian National Data Service Projects (ANDS);
\item the \gaia-ESO Public Spectroscopic Survey \citep[GES,][]{GES_final_release_paper_1,GES_final_release_paper_2}. The \gaia-ESO Survey is based on data products from observations made with ESO Telescopes at the La Silla Paranal Observatory under programme ID 188.B-3002. Public data releases are available through the \href{https://www.gaia-eso.eu/data-products/public-data-releases}{ESO Science Portal}. The project has received funding from the Leverhulme Trust (project RPG-2012-541), the European Research Council (project ERC-2012-AdG 320360-Gaia-ESO-MW), and the Istituto Nazionale di Astrofisica, INAF (2012: CRA 1.05.01.09.16; 2013: CRA 1.05.06.02.07).
\end{itemize}

The GBOT programme  uses observations collected at (i) the European Organisation for Astronomical Research in the Southern Hemisphere (ESO) with the VLT Survey Telescope (VST), under ESO programmes
092.B-0165,
093.B-0236,
094.B-0181,
095.B-0046,
096.B-0162,
097.B-0304,
098.B-0030,
099.B-0034,
0100.B-0131,
0101.B-0156,
0102.B-0174, and
0103.B-0165;
%
%
and (ii) the Liverpool Telescope, which is operated on the island of La Palma by Liverpool John Moores University in the Spanish Observatorio del Roque de los Muchachos of the Instituto de Astrof\'{\i}sica de Canarias with financial support from the United Kingdom Science and Technology Facilities Council, and (iii) telescopes of the Las Cumbres Observatory Global Telescope Network.
\end{appendix}
\end{document}